\documentclass{article}

\usepackage[utf8]{inputenc}
\usepackage{amsmath,amssymb}
\usepackage{cite}
\usepackage{graphicx}
\usepackage{epstopdf}
\usepackage{longtable}
\usepackage{caption}
\usepackage{colortbl}

\date{}

\newcommand{\Q}{\mathbb Q}
\newcommand{\N}{\mathbb{N}}
\newtheorem{definition}{Def\,inition}

\newcolumntype{H}{>{\setbox0=\hbox\bgroup}c<{\egroup}@{}}

\begin{document}
\vspace*{0.35in}

\begin{flushleft}
{\Large
\textbf\newline{Protein Repeats from First Principles}
}
\newline
\\
Pablo Turjanski\textsuperscript{1},
R. Gonzalo Parra\textsuperscript{2},
Roc\'io Espada\textsuperscript{2},
Ver\'onica Becher\textsuperscript{1},
Diego U. Ferreiro\textsuperscript{2,*}

\bigskip
\bf{1} Departamento de Computaci\'on, Facultad de Ciencias Exactas y Naturales, Universidad de Buenos Aires, Buenos Aires, Argentina
\\
\bf{2} Protein Physiology Lab, Departamento de Qu\'imica Biol\'ogica, Facultad de Ciencias Exactas y Naturales, Universidad de Buenos Aires-CONICET-IQUIBICEN, Buenos Aires, Argentina
\\
\bigskip

* ferreiro@qb.fcen.uba.ar

\end{flushleft}
\section*{Abstract}
Some natural proteins display recurrent structural patterns.
Despite being highly similar at the tertiary structure level, 
repetitions within a single repeat protein can be extremely variable at the sequence level.
We propose a mathematical definition of a repeat and investigate the occurrences of these in different protein families. 
We found that long stretches of perfect repetitions are infrequent in individual natural proteins, even for those which are known to fold into structures of recurrent structural motifs. 
We found that natural repeat proteins are indeed repetitive in their families, exhibiting abundant stretches of  6 amino acids  or longer that are perfect repetitions in the reference family. We provide a systematic quantification for this repetitiveness. We show that this form of repetitiveness is not exclusive of repeat proteins, but also occurs in globular domains. A by-product of this work is a fast classifier of proteins into families, which yields likelihood value about a given protein belonging to a given family.

\section*{Introduction}
Natural repeat proteins are coded with tandem copies of similar 
amino acid stretches. These molecules are broadly classified according to the length of the minimal repeating unit \cite{Kajava}. Short repetitions of up to five residues usually form fibrillar structures, while repetitions longer than about 60 residues frequently fold as independent globular domains. There is a class of repetitive proteins that lays in between these for which folding of the repeating units is  coupled and domains are not obvious to define \cite{Espada,Parra}. Despite being highly similar at the tertiary structure level, repeats within a single protein or  in different members of a protein family can be extremely variable at the sequence level \cite{Espada2}, complicating the detection and classification of repeats \cite{Kajava}.

There are many methods to identify repeats in sequences. 
Some are based on the self-alignment of the primary structure \cite{Luo}
and others implement spectral analysis of pseudo-chemical characteristics
of the amino acids \cite{Marsella}. Since the same structural motif can
be encoded by sequences that seem completely unrelated, it is not surprising
that alignment-based methods fail to infer true structural repeats.
The solutions to find inexact repeats in sequences  \cite{Crochemore,Gusfield}
include alphabet replacements using scoring matrices, 
sophisticated notions of sequence similarity based on an allowed percentage of mismatches, 
and elaborated mathematical representations such as Hidden Markov Models.
To a very large extent these solutions have been satisfactory.
However, these methods rely on the fine-tuning of different
parameters in order to account for the inexactness of repeats
(thresholds for alphabet scoring matrices, allowed percentage of mismatches, 
e-values for Hidden Markov Models and others).
The definition of what constitutes or not a hit for the model remains 
subject to some threshold definition.

In this work we turned to ``first principles'', starting with a mathematical definition of biological repeats and we developed a repeat finding method with no adjustable parameters. We use the concept of maximality and maximal repetition (MR). In the context of a protein sequence, a MR is a block of amino acids that occurs two or more times and any of its extensions occurs fewer times. In case a long block in a protein sequence is equal to another, except for one letter, then there will be two repetitions, one to the left and one to the right of that single letter. It is well known that long stretches of perfect repeats are infrequent in natural proteins, even in those that 
 fold into structures of recurrent structural motifs. However, we observe that a large portion of a protein sequence can 
be described by short stretches of amino acids that occur  in other members 
of a protein family. Thus, a protein family operates as a catalogue of all the possible variations that a  block can adopt in any of its members. We quantify the repetitions we measure how well a given sequence is covered by the repetitions occurring in its family.  The method is implemented efficiently by an algorithm with a O($n \, log \, n$) computational complexity, where $n$ is the size of the protein sequence being tested. From this quantification we directly obtain a way to decide if one family is more repetitive than another. In addition, this quantification allows us to derive a measure of likelihood for a given sequence to belong to a given family.

\subsection*{Notation and preliminary definitions} 
\newcommand{\M}{\+M}
Let $\+A$ be an alphabet, which is  a finite set of symbols.
We consider sequences of symbols in $\+A$.
The length of a sequence $s$ is denoted by $|s|$. 
We address the positions of a sequence $s$ by counting from $1$ to $|s|$. 
With $s[i..j]$  we denote the sequence that starts in position $i$ and ends in position $j$ in $s$.
If $i$ or $j$ are out of range then $s[i..j]$ is equal to the empty sequence. 
We say $u$ occurs in  $s$ if $u = s[i..j]$ for some $i,j$.
In case $s$ starts with $u$ we say  that $s$ is an extension of $u$. 

\begin{definition}[Gusfield \cite{Gusfield}] \label{def:maxrep}
A {\em maximal repeat} (MR) is a sequence that occurs more than once in $w$, and each of its extensions occurs fewer times.
We write  $\M(s, n)$ to denote the set of MRs of lengths greater than or equal to $n$, 
that occur in the sequence $s$. 
\end{definition}

\begin{figure}[ht!]
 	\includegraphics[width=1\linewidth]{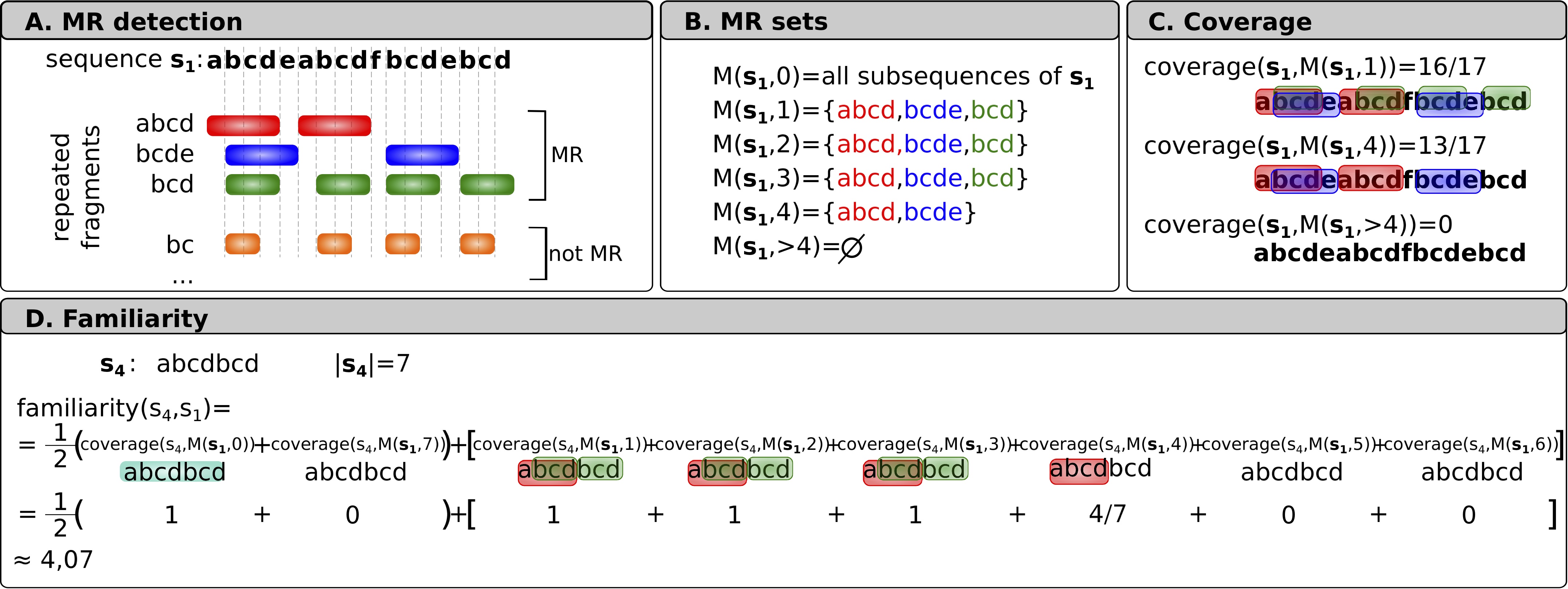}
	\caption{{\bf Scheme of the procedure to obtain \em{familiarity} \bf function value.}
	A) Maximal repeats (MR) are computed for input sequence. B) MR sets are filtered by the minimum MR length. C) MR sets are overlapped to input sequence and coverage is calculated. D) $familiarity$ is computed based on coverage at every length. }   
        \label{Fig1}
\end{figure}

The set of MRs of  $s_1=abcdeabcdfbcdebcd$ is  $\{ abcd,\  bcde, bcd \}$. 
Observe  that $abcd$ and $bcde$ are the longest MRs, occurring twice. 
But $bcd$ is also a MR because it occurs four times in $s_1$, 
and every extension of $bcd$ occurs fewer times. 
On the contrary, $bc$ is not a  MR because 
both $bc$ and $bcd$ occur four times, 
contradicting the condition that the extension must occur fewer times 
(see Fig. \ref{Fig1}A).
The set of MRs of $s_2=aaaa$ is $\{ aaa,\  aa, a \}$ where $aaa$ is a MR occurring twice, 
$aa$ occurs three times and $a$ four times.
The set of MRs of $s_3=ab$ is empty.

From the given examples is easy to see that MR occurrences can be nested and overlapping.
The MRs in any given sequence $s$ can have lengths between 1 and a maximum of $|s|$-1 (this maximum is reached only when the sequence is a chain of the same letter, as  $aaaa$).
The total number of different MR patterns in a sequence of length $|s|$ is at most $|s|$ because there is at most one different MR starting at each position.

\begin{definition}
Let $S$ be a set of $n$ sequences over the alphabet $\+A$, $S=\{s_1,s_2, \ldots s_n\}$.
The set of MRs in $S$ is the set of MRs of the sequence obtained by concatenation of 
all sequences in $S$, interleaved with pairwise different symbols 
$\$_1, \ldots, \$_{n-1}$ that are not in $\+A$. 
Thus, the set of MRs in $S$ is the set of MRs in $s_1\$_1s_2\$_2\ldots \$_{n-1}s_n$. 
\end{definition}

Since each $\$_i$, for $i=1,2, \ldots n-1$, occurs only once in the concatenated sequence there will be no  MRs containing them. 
Since the symbols $\$_i$, for $i=1,2, \ldots, n-1$, are pairwise different, the set of MRs is invariant respect to the order in which we concatenate the sequences $s_1, s_2, \ldots, s_n$. 
Concatenation of any permutation of the sequences $s_1, \ldots, s_n$  
produces the same set of MRs.

Observe that finding the set of MRs in a set of sequences requires more than treating them individually. If $s$ and $t$ are two sequences and \$ is a symbol not occurring in $s$ nor in $t$, the  MRs in $w = s\$t$ may be different from getting the individual MRs and take their union, because a repeat in $w$ may 
occur only once in $s$ and only once in $t$.

\paragraph{The familiarity function}
We define the familiarity function that measures how much of a given protein sequence is covered by MRs from certain family. The greater the familiarity value, the most likely the protein sequence to belong to the family. We first introduce the classical notion of {\em coverage} of a sequence by a set of MRs, which measures the number of positions in the sequence that are covered by the MRs in the set.
We write $\+A^*$  for the set of all sequences over~$\+A$, and 
$\+P(\+A^*)$ the set of all  the parts  $\+A^*$, 
which represents the collection of all the different sets of sequences over~$\+A$.
As usual, we write $\N$ and $\Q$ for the set of natural and rational numbers, respectively.

\begin{definition}
The function $coverage:\+A^* \times \+P(\+A^*) \to \Q$ is such that 
for any sequence $s$ and any set of sequences $R$
\noindent

\begin{equation}
coverage(s, R) =
\frac{\#\{j:\exists i \in \N , \exists r \in R,\ s[i..i+|r|-1] = r \} }{ |s|}.
\end{equation}

Thus, $coverage(s, R)$ is a rational number between $0$ and $1$.
\end{definition}
For example, for $s_1=abcdeabcdfbcdebcd$ and 
$R=\M(s_1,1)=\{abcd, bcde, bcd\}$ we have $coverage(s_1,R)=16/17\approx0.94$ (see Fig. \ref{Fig1}C).

The {\em familiarity} function measures how much of a sequence is covered by 
a set of MRs that occur in a family.

\begin{definition}
The $familiarity$ function $:\+A^* \times \+A^* \to \Q$ is defined as follows.
For any sequence $s$ and any sequence $t$, 

\begin{equation}
familiarity(s,t)=
\frac{coverage(s,\M(t,0))+ coverage(s,\M(t,|s|))}{2} 
 + \sum_{i=1}^{|s|-1} coverage(s,\M(t,i))
\end{equation}

\end{definition}
Note that $familiarity(s,t)$ uses $\M(t,0)$ which, by definition, gives all the blocks of the sequence $t$.
Thus, for every sequence $s$ and $t$ the function $familiarity(s,t)$ is a number between $0$ and $|s|$.
For example, the $familiarity(s_4,s_1)$ of $s_4=abcdbcd$ and $s_1=abcdeabcdfbcdebcd$ is around $4.07$ because the set of MRs of $s_1$ is $\{abcd,bcde,bcd\}$, and then $familiarity(s_4,s_1)=\frac{1 + 0}{2} + 1 + 1 + 1 + \frac{4}{7} + 0 + 0 \approx 4.07$ (see Fig. \ref{Fig1}D).

If the $familiarity$ function is evaluated with the same sequence in the two arguments, $familiarity(s,s)$ the result tells how much of the sequence $s$ is covered by its own MRs. For example, the $familiarity(s_5,s_5)$ of $s_5=abcabca$ is $4.5$ because the set of MRs of $s_5$ is $\{a,abca\}$, and then $familiarity(s_5,s_5)=$ $\frac{1 + 0}{2} + 1 + 1 + 1 + 1 + 0 + 0 = 4.5$.
In the case of $s_6=aaaaaaa$ the $familiarity(s_6,s_6)$  is $6.5$ because the set of MRs of $s_6$ is $\{a,aa,aaa,aaaa,aaaaa,aaaaaa\}$, and then $familiarity(s_6,s_6)=$ $\frac{1 + 0}{2} + 1 + 1 + 1 + 1 + 1 + 1 = 6.5$. In these examples, $s_6$ reaches a higher coverage than $s_5$ when using MRs internal to each of them.

For a given set of sequences, let $t$ be the concatenation of its elements separated by pairwise different symbols. Then,
$
familiarity(s, t)
$
indicates how much of the sequence $s$ coincides with  the MRs in $t$. 
For example, the $familiarity(s_5,t_1)$ of $s_5=abcabca$ and $t_1$=$aa\$_1ab\$_2 adddd\$_3bca$ is around $1.21$ because
the set of MRs of $t_1$ is $\{a,b,d,dd,ddd\}$, and then
 $familiarity(s_5,t_1)=\frac{1 + 0}{2} + \frac{5}{7} + 0 + 0 + 0 + 0 + 0 \approx 1.21$. 
Hereafter we will just use the name of a family in the second argument of the {\em familiarity} function, 
to denote the concatenation of all the sequences present in that family, separated by pairwise different symbols.

\section*{Results and Discussion}

\subsection*{Maximal repeats inside protein sequences}  
\begin{figure}[ht!]
 	\includegraphics[width=1\linewidth]{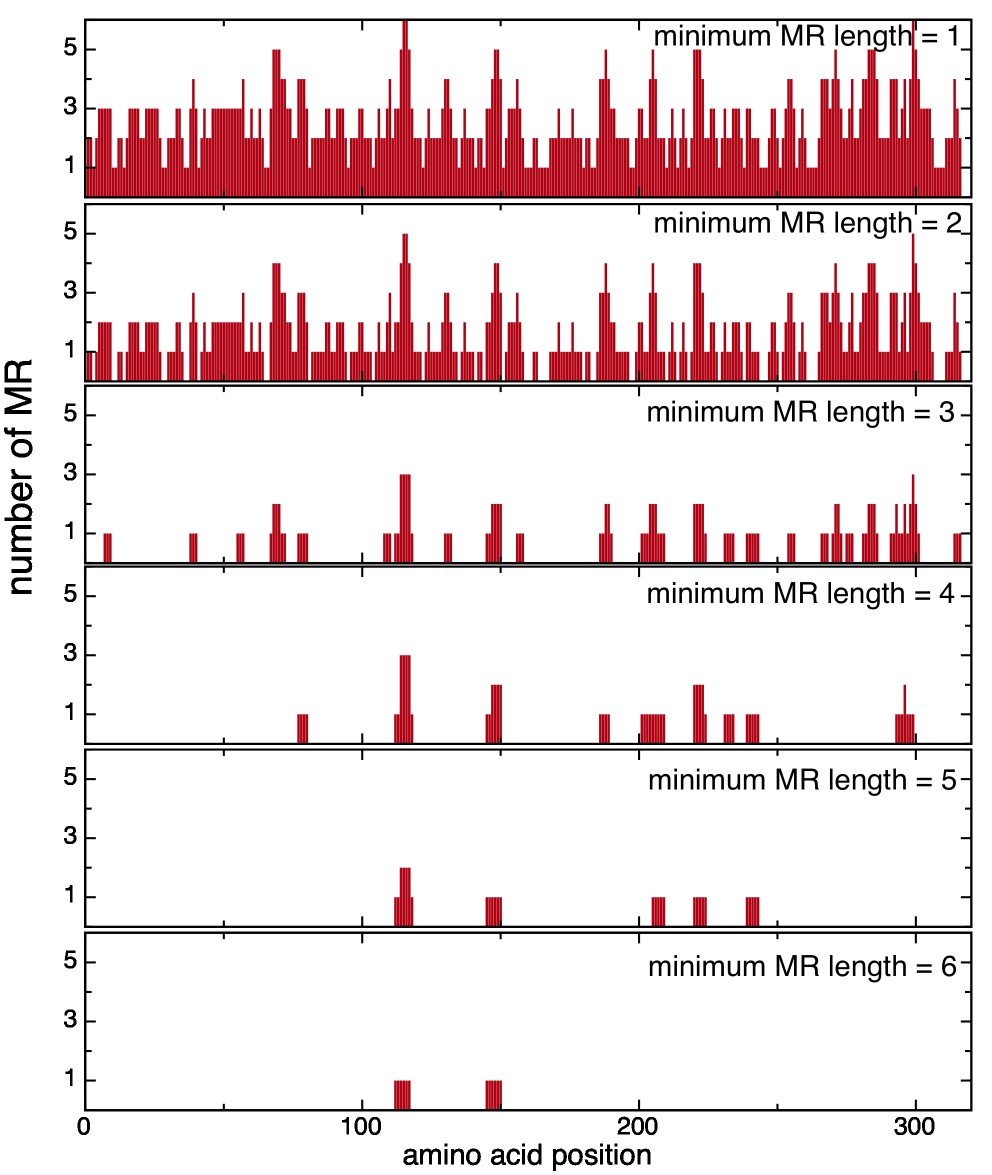}
	\caption{{\bf Number of maximal repeats (MR) that affect each position on a trial sequence.} 
	The I$\kappa$B$\alpha$ protein sequence was used as input, the MR set were computed by $\M($I$\kappa$B$\alpha,i)$ with $i=1\dots6$. The panels show the counts per position for the different MR sets sorted by minimum length.}
       \label{Fig2}
\end{figure}

 Since some proteins contain clear repetitive motifs in structure, 
we wondered  how much of that repetitiveness is maintained at the sequence level. 
We analyzed the occurrence of exact repetitions on members of the Ankyrin repeat protein family (ANK$_t$), for which many structures have been solved. Ankyrins constitute the most abundant class of natural repeat proteins, and have been extensively studied. 
We computed the maximal repeats (MRs) inside each protein, 
for all possible lengths (from a minimum length of 1 to the maximum possible, 
the length of the sequence minus 1). 
Fig. \ref{Fig2} shows the coverage given by MRs inside $s$=$I \kappa B \alpha ($\textit{Uniprot ID: P25963}), a member of the ANK$_t$, for different minimum MRs lengths ($\M($I$\kappa$B$\alpha,i)$ for $i=1,\dots,6$). 
We find that this protein has 102 MRs distributed as follows: 
20 MRs of length 1, 65 MRs of length 2, 11 MRs of length 3, 3 MRs of length 4, 2 MRs of length 5 
and only one MR of length 6. 
The detected MRs are not evenly distributed along the sequence 
but clustered at specific positions. In most cases the shorter MRs are nested within longer MRs. 
Moreover, several MRs occur in the same parts of the sequence. These are overlapping occurrences of MRs. 

\begin{figure}[ht!]
  	 \includegraphics[width=1\linewidth]{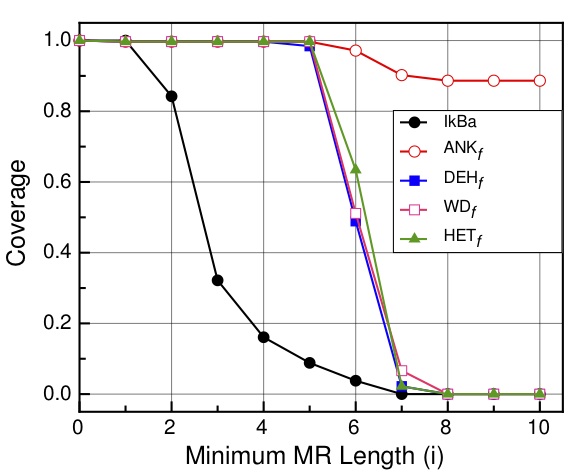}
         \caption{{\bf Coverage of a trial sequence of I$\kappa$B$\alpha$ using different MR sets.}
	 The maximal repeats (MR) sets were computed from the sequence itself (black) or using groups of sequences derived from distinct families. Values come from applying $Coverage($I$\kappa$B$\alpha,\M(t,i))$ function for $t=$I$\kappa$B$\alpha$, ANK$_f$, DEH$_f$ WD$_f$, HET$_f$ and $i=0,\,...\,,10$ }
\label{Fig3}
\end{figure}

We analyzed the coverage of the primary structure of the protein I$\kappa$B$\alpha$ 
using sets of MRs of increasing minimum length (Fig. \ref{Fig3}, black dots). 
Trivially, the coverage is maximum when MRs of length $1$ are considered, 
because in general every amino acid occurs at least twice inside the protein and then, every position in the protein is covered 
by some MR of length~$1$. 
The coverage is reduced as the minimum MR length is increased, 
reaching $0$ for the values of $i=7,\dots, |s|$, 
as there are no exact repetitions larger or equal than 7 residues. 
Coverage values for all the members of the ANK$_t$ were calculated for 
maximal lengths $i=0,\dots,10$ (see Table in Table S\ref{S1_Table}).
For each ANK$_t$ protein, the set $M(s,1)$ produces almost full coverage (the coverage function is $\simeq1$). 
However, the set of MRs of length $i$ decays rapidly as $i$ 
increases, and very soon the set of MRs becomes empty. 
The set of MRs of lengths $i>=6$ contrasts with the MRs that can be found in structures, where no sequence information is taken into account \cite{Parra}. 
In general, most of the Ankyrin repeat proteins (ANKs) analyzed in this work, are almost entirely covered by structural repeats.

There is a subtype of ANK proteins, which are synthetic constructs that are composed of (nearly) identical repetitions, for which, as expected, we detected long MRs in sequence. 
The molecules, such as DARPINs, OR264, OR266 and NRC, have a much larger 
coverage than natural ANKs, realized by their  long perfect repeats
(which are directly connected to the  construction methods)
\cite{Mosavi,Binz} (see Table in Table S\ref{S1_Table}).

The results we obtained for ANK$_t$ were contrasted with results of two other protein families. 
WD-40 proteins (WD$_t$) are non-solenoid repetitive proteins with mainly $\beta$ composition, that fold into a globular-like $\beta$-propeller fold. 
For these proteins, the distribution of MR lengths is similar to that of ANKs, 
with infrequent exact repeats larger than 6 residues 
(see Table in \ref{S3_Table}). 
Additionally we tested our program in a non-repetitive globular scaffold 
using members from the Dehalogenase family (DEH$_t$). 
Results are shown in Table in \ref{S2_Table}. 
In general, the $coverage$ function applied to DEH$_t$ proteins goes to zero for  lower $i$ values than in the repeat protein families, indicating that the MRs in DEH$_t$ proteins are shorter than the MRs in repeat proteins.
The low coverage of the sequences may be consequence of their divergence during evolutionary time scales.
Most of existing methods for repeat detection in protein sequences partially fail when proteins are too divergent respect a consensus sequence. 
However, the identification  of the individual occurrences of repeats is simple to observe at the structural level. 
The higher conservation of the repetitive patterns in structures has 
been recently exploited to visually classify and  annotate these kind of proteins \cite{DiDomenico}. 
Since protein sequences encode protein structures, 
we believe there must be a way to unravel the sequence repetitiveness despite 
the dissimilarity among the repeats.

\subsection*{Maximal repeats in families}
As we have seen previously, long stretches of perfect repeats are infrequent in natural proteins, 
even for those which are known to fold into structures of recurrent structural motifs. 
Sequence-wise, repeats are known to be imperfect. Unfortunately, the methods that assume repeats to be degenerated fail to make a complete detection.
Also these methods do not allow to conclude if some individual motifs actually occur or not. 
For instance, in ANKs, there are some specific sub-motifs that are characteristic of the family when looking at the statistical profile of ANK repeats, as a TPLH motif and variations of it; however, when looking at particular individual sequences it is hard to say whether they correspond to ANK instances or not. 
All possible variations of typical blocks should be represented in at least on member of the family. Sequence statistical profiles, usually assume that positions are independent. Therefore, when combining different amino acids at adjacent positions, blocks that are not representative of the family can be constructed, since natural covariations are not taken into account. The opposite, i.e natural occurring blocks that are a consequence of combinations of amino acids with low frequencies may not be detected as part of the motif. We overcome this problem by looking for natural occurring blocks in members of the family. This additionally solves the problem of position independence since these are implicitly used in the short repetitions.

\begin{figure}[ht!]
	\includegraphics[width=1\linewidth]{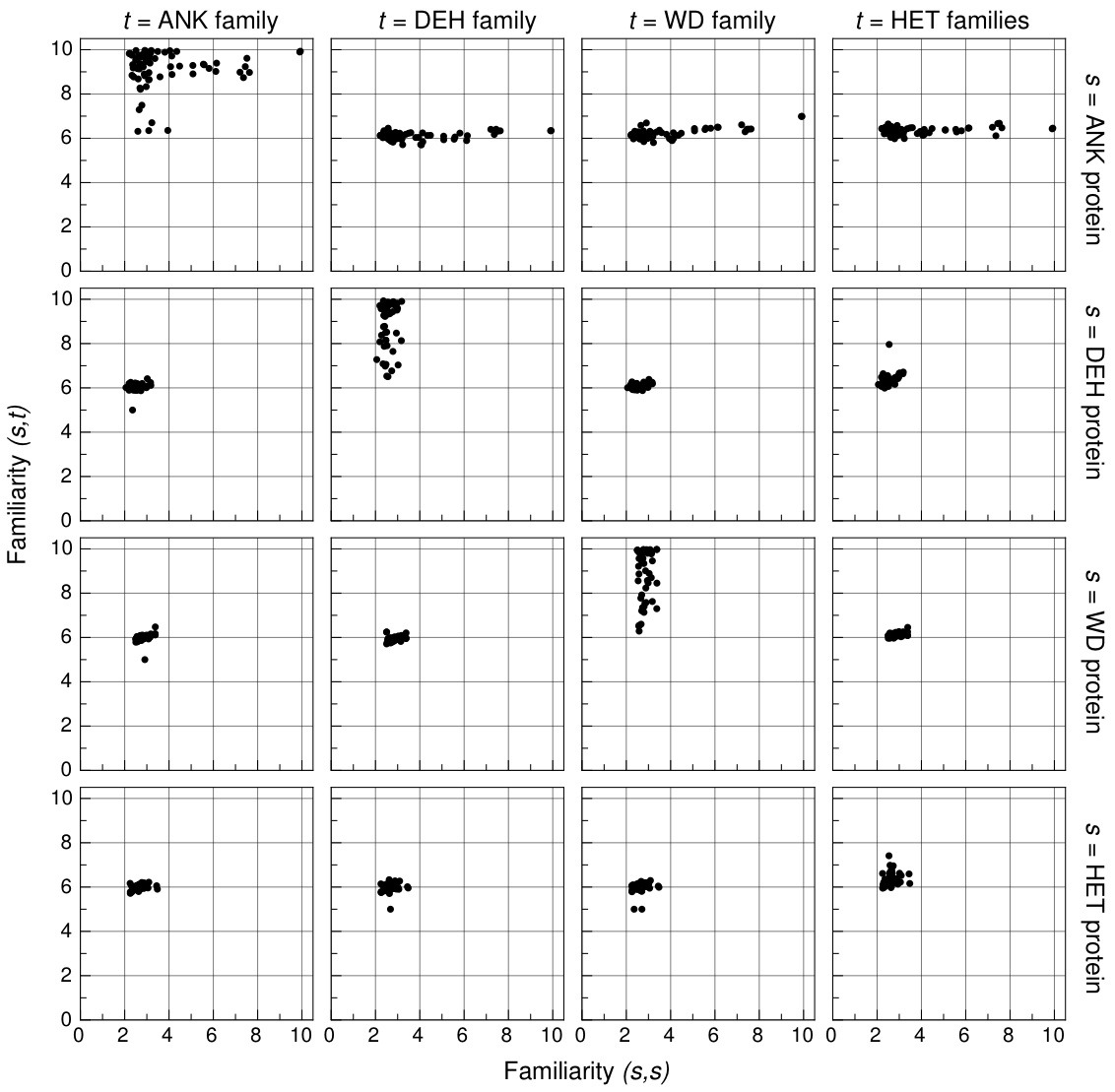}
       \caption{{\bf \em{$familiarity$} \bf of 223 trial protein sequences was calculated using maximal repeats (MR) sets derived from distinct sequences.}
       The x-axis denotes the  $familiarity(s,s)$ of the proteins, $s$, with MR set calculated from the sequence itself, the y-axis shows the $familiarity(s,t)$ calculated with MR sets derived from distinct families, $t$.}        
	\label{Fig4}  
\end{figure}

Given a sequence $s$ and a family $f$ our method consists in finding the repetitions in the family $f$ that have some occurrence in the sequence $s$.
We first compute  
the sets $\M(t,i)$, where $t$ is the concatenation of the sequences in $f$ separated by pairwise different symbols, 
for all the possible values of values of  $i$, namely $i$ goes from 1 to $|s|$.
We then compute the coverage made by the elements of $\M(t,i)$
on the sequence $s$ using the $coverage(s,\M(t,i))$ function. 
As example, Fig. \ref{Fig3} presents the coverage of the I$\kappa$B$\alpha$ protein 
considering sets of MRs from different families. 
The coverage was calculated by the $coverage$(I$\kappa$B$\alpha$, $\M(t,i)$) function, 
using $t=$ I$\kappa$B$\alpha$ alone, ANK$_f$, DEH$_f$, WD$_f$ and HET$_f$ datasets and $i=0,\dots,10$. 
The HET$_f$ dataset is a selection of proteins from different families. 
We observe that, as the minimum MR length increases above $i=3$ the 
$coverage$(I$\kappa$B$\alpha$, $\M($I$\kappa$B$\alpha,i)$) decays under 0.02 (black line), while the coverage remains close to 1 for MRs detected in larger datasets up to $i= 6$. 
The coverage only keeps significantly high for longer MRs when using the set of MRs obtained from the ANK family, to which the protein belongs.
With these results, we computed the $familiarity($I$\kappa$B$\alpha,t)$ function 
for $t=$ I$\kappa$B$\alpha$ alone, ANK$_f$, DEH$_f$, WD$_f$ and HET$_f$ datasets. 
Although the definition of $familiarity$ requires the values of $coverage(s,\M(t,i))$ for each  $i$ in $  [0..|s|]$,
in all the cases we analyzed it was  enough to consider $i$ just  in $[0..10]$, because the coverage for larger values of $i$ is negligible. 
Hereafter we consider the $familiarity$ function  with lengths $i \in [0..10]$. 
The maximum coverage is obtained for $familiarity$(I$\kappa$B$\alpha$, ANK$_f$)$=$9.57571. 
$familiarity$ function applied to I$\kappa$B$\alpha$ together with other families have values less than 6.15
(see Table in \ref{S4_Table}, Uniprot ID = P25963). 
This function  indicates that I$\kappa$B$\alpha$ belongs to the ANK$_f$ family. To verify if this hypothesis can be generalized,
we also applied the previous $familiarity$ function to each of the 223 test sequences for which the protein structure 
is known. These test sequences include 73 ANKs (ANK$_t$), 50 DEHs (DEH$_t$), 50 WD-40s (WD$_t$) and 
50 randomly selected proteins that do not belong to the previous families (HET$_t$). 
For more details on how these sets were constructed see the section on Materials and Methods.
Values of this function are shown in Table in \ref{S4_Table}. 
A global visualization of the values can be seen in Fig. \ref{Fig4}.
As for the I$\kappa$B$\alpha$ protein, we observe that 
the familiarity function for the sequence $s$ itself lays in between $2$ and $4$, 
except for designed ANKs that have higher values. 
These low values are a consequence of the low number of exact repeats 
in their sequence composition, regardless they belong to a repetitive or globular family. 
Even when repetitiveness is evident at the structural level, 
sequences from repetitive and globular proteins are indistinguishable in terms
 of the covering done by their own MRs. 
On the contrary, we observed a clear difference for the familiarity values 
 for these sequences when evaluated in the context of  families. 
For all proteins belonging to the ANK$_t$, WD$_t$ and DEH$_t$ 
the familiarity values are in a range from $6$ to $10$ when evaluated in the context of their own family,
but drop to values around $6$ when evaluated in the context of a  family different to its own.
One interpretation of these results is that protein families constitute ensembles where each of its members is composed of perfect repeats that are present in other members of the ensemble.

To our surprise, we were not able to see differences between the familiarity values of repeat families and  globular families.
Our hypothesis is that this high familiarity value is a common feature of protein
 families  that are equilibrated ensembles whose members are mostly composed of exact repetitions ranging from dipeptides to decapeptides. 
This also suggests  that natural proteins are built up from fragments longer than dipeptides but shorter than decapeptides, 
in line with the general ideas implemented by `fragment assembly' of synthetic proteins \cite{Moult}.

There are however, some notable exceptions in the results of familiarity values in our experiments. 
In the case of ANK$_t$, protein $P14585$ is composed of more than 1400 residues, 
but its ANK region encompasses only about 200 amino acids. 
As a consequence of this, the coverage (and familiarity) obtained for this sequence in the context of the ANK$_f$ 
set does not display values significantly higher than for the other sets corresponding to foreign families. 
 
Other exceptional cases within the ANK$_t$ group of sequences that are not well explained 
by MRs found in the ANK$_f$ set are proteins that fold into ANK-like structures 
but strongly differ from  the rest of the family in their sequence patterns. 
These cases correspond to sequences $Q5ZSV0$ and $Q5ZXN6$ from \textit{Legionella sp}, 
sequence $Q978J0$ from \textit{Thermoplasma volcanium}, sequence $O22265$ which is the 
only protein from a plant in the dataset, and sequence $Q6IV60$ which is a viral protein. 
Except for the plant protein, all these cases are non-eukaryote proteins. The ANK motif is known to be particularly enriched in eukaryotes and within specific eukaryote pathogens (including bacteria and viruses) that use ANK-like proteins to mimic their host counterparts and proceed with the infectious processes \cite{Voth}. The origin of non-eukaryote ANK-like proteins has been discussed with no consensus about whether  they correspond to horizontally transferred molecules with subsequent divergent evolution, or  they originated by convergent evolutionary processes. 
\subsection*{How are maximal repeats distributed in the families?} 

\begin{figure}[ht!]
 	\includegraphics[width=1\linewidth]{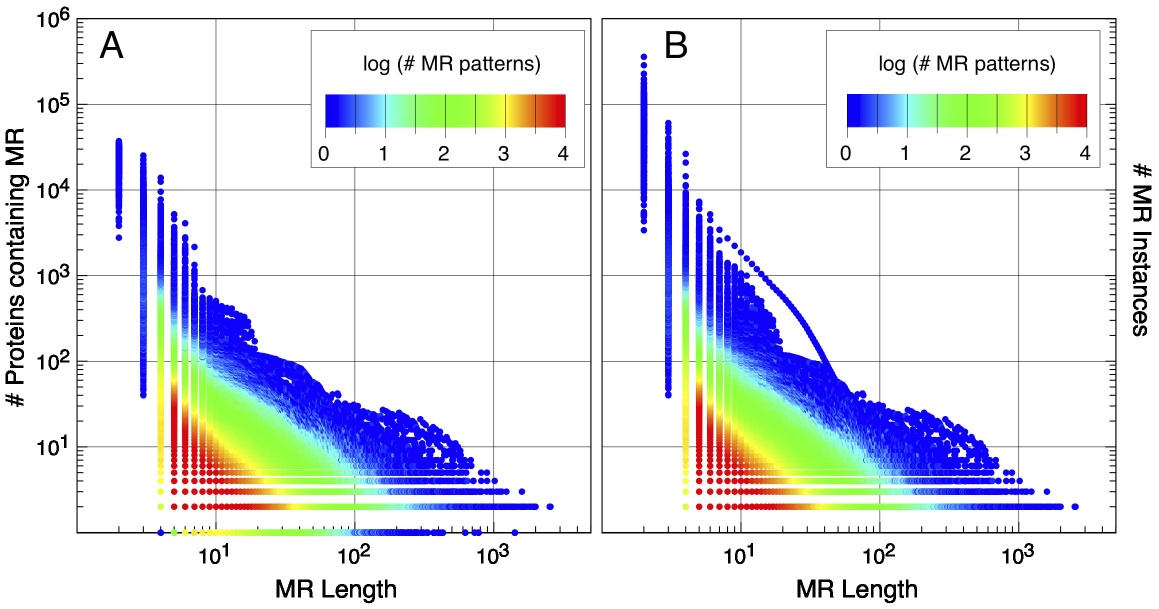}
	\caption{{\bf The sequences of the ankyrin family (ANK$_f$) were used to calculate the maximal repeats (MR) set.}
	The distribution of the millions of MR found on the whole set is shown according to the length of the pattern. A) Number of different proteins that contain the MR pattern. B) Number of times each MR pattern is present in the whole dataset. The colorscale denotes the number of different MR patterns that occur at a particular coordinate.}   
       \label{Fig5}
\end{figure}

For each family dataset (ANK$_f$, DEH$_f$, WD$_f$ and HET$_f$) we evaluated how its MRs are distributed within the proteins members of the family. 
We counted how many proteins in the family contain each of the MRs that are found in that family (Fig. \ref{Fig5}A shows the case of ANK$_f$, and Fig. S1
 \, shows the case for the remaining datasets). 
We observe that there is a large number of short MRs occurring in many different proteins (e.g. ``TP'' appears over 85\% of ANK$_f$ proteins), and a small number of long MRs occurring in just a few different proteins (e.g. ``GNPFTPLHCAVINDHE'' appears only in two proteins from ANK$_f$). The longest MR sequence (2,563 residues) has only two instances and appears in two very similar proteins (F1MVI7 and G3MYJ1).

Considering that MRs with lengths $i>=6$ generally are only found in the family's own proteins, we focused on MRs corresponding to those $i$ values. 
ANK$_f$ has a total of $38,051$ proteins but MRs larger than $6$ residues do not appear in all of them (the most popular MR of length $6$ appears in $4,085$ proteins, but in average MRs of this length appear in less than $1,000$ proteins). 
Thus, there is no block larger or equal than six, common to the whole family.
The coverage of each member of the ANK family by MRs in ANK$_f$ comes from 
many different proteins. 
In this sense, the ANK proteins seem to be a mosaic of exact MRs 
that are spread along their whole sequence.
The number of mismatches in individual sequences of a given family, 
found by pairwise alignments inside each sequence, 
vanishes dramatically if we consider the repeating blocks in other proteins from the family.

In ANK$_f$ we observe that, starting from MRs of lengths $i>=6$, 
Figs. \ref{Fig5}A and \ref{Fig5}B are very similar. This shows that MRs occur at most once within each protein sequence. It also proves that long stretches of perfect repeats are infrequent in natural proteins, even for specific members that are known to fold into structures of recurrent structural motifs. However, there is a collection of recurring MRs that are spread along the members of a family than can be used to (partially) reconstruct any given sequence of the family. 

The same analysis over other families (DEH$_f$ and WD$_f$) shows similar results to those for ANK$_f$ (see Fig. S1
). We also considered a dataset of sequences constructed by scrambling the proteins of a given family 
(this dataset constructed by scrambling the amino acids of each sequence from HET$_f$ dataset). The result differs completely from the results for actual families 
(Fig. S1
). 
This is expected because scrambled sequences have less and shorter repeats. 
We made the same analysis for the HET$_f$ dataset (Fig. S1
). Although the HET$_f$ has longer MRs than the scrambled protein family dataset, these MRs are notoriously shorter than the MRs in the ANK$_f$, DEH$_f$ and WD$_f$ families. 
This experiment gives evidence for the difference between sets of sequences 
that constitute an actual protein family and sets of proteins which do not constitute a family, as compared to sequences that do not correspond to actual proteins.

\subsection*{Towards a catalogue of repeats}
We computed the set of MRs of length $6$ or longer from the ANK$_f$ dataset, this is $\M(ANK_f,6)$. 
The minimum length value of $6$ was selected to compare what we observe in small structural repetitions 
with the tiling methodology \cite{Parra}. As a result we obtained $4,390,695$ MRs with a length of $6$ residues  
which exponentially decreases as the MRs length increases. 
The most frequent MRs, for instance  TPLHLA and GADVNA (and their variants), 
coincide with the most popular motifs in the ANK HMM profile (Pfam ID: PF00023). 
We computed the proportion between instances of the MRs and the number of proteins containing them. 
The most well known motifs, have a proportion close to $1$. 
However, we found several other motifs to be quite popular, as LISHGA, GHLDVV and  ELLISH.
They have a higher proportion between instances of the MRs and the number of proteins containing them 
(between 2 and 3), and they are conserved along repeat domains in the ANK$_f$ dataset.
The particularity of these motifs is that their occurrences are not evident when visually 
observing the sequence logo representation for the ANK HMM profile, because
they are composed of highly frequent amino acids at some positions 
and infrequent amino acids in others. 
The identified motifs respect the short length covariation in between positions, 
which is not taken into account in HMMs, in which the positions are assumed to be independent. 
Consequently, strategies like scanning sequences with HMM profiles need to apply a 
threshold to accept or not a subsequence as a hit. This can lead to spurious amino 
acid combinations producing false positives or false negative results. 
Using short exact sequences in order to look for MRs, considers implicitly the natural covariation among the residues that constitute them and at the same time allows us to avoid the use of thresholds.

\subsection*{Concluding Remarks}

We posed the question: \textit{How repetitive are natural repeat-proteins?}
We committed to a mathematical definition of a repeat and found that long stretches of perfect repeats are infrequent in natural proteins, 
even for those which are known to fold into structures of recurrent structural motifs. However, we found that repeat proteins have  abundant stretches of $6$ amino acids or longer that are perfect repetitions in the reference family. We provided a systematic quantification for this repetitiveness.

Our solution finds all the  maximal perfect repetitions, using no adjustable parameters. We use a reference family of protein sequences, that operates as a catalogue of all the possible variations that  repeating blocks can adopt. We show that a large portion of each protein sequence can be described 
by short stretches of amino acids occurring in members of the reference family. 
Thus, each family determines an expected covering of its sequences by family repeats.
This yields a measure of likelihood for any sequence to belong to a given family, quantified by the $familiarity$ function. This function actually provides a classification method for proteins in families. The method could be used  as a guiding tool  in the design of synthetic proteins, establishing a minimum and a maximum value of a candidate sequence in relation to existing families.

The familiarity function can be implemented with an algorithm whose computational complexity is  O($n \, log \, n$),  where $n$ is the size of the protein sequence plus the size of the family dataset. This allows to compute the classification very efficiently. 

This work is rooted in an exhaustive analysis of three natural protein families and two control families, taken as examples. 
The study can be extended to cover the complete protein universe or a substantial part of it. 
Moreover, the approach does not require a detailed curation of the sequences present in the families.
We have limited our current work to the  identification of   MRs in families   and to the computation of the familiarity function. 
Detailed statistical work remains to be done on MRs in families, such as the average distance 
between different occurrence of MRs inside the same protein sequence, the number of different MRs 
per length in each protein sequence. We also suggest to identify the subset of overlapping MRs 
(and the size of the overlap), the subset of  non-overlapping MRs, the subsets of MRs that can be placed one after another, 
and the subset of MRs that exclude the occurrence of others.
These statistics may yield relations between  maximal repeats with   some known functional features
and to some further  conditions for the construction of synthetic proteins.

\section*{Materials and Methods}

\subsection*{Protein family datasets} 
\paragraph{Ankyrin repeat protein (ANK$_f$) and WD40 repeat protein (WD$_f$) families datasets.} 
From Uniprot Uniref0.9 we run hmmsearch from the hmmer suit for a specific HMM family taken from PFAM. 
Included only  sequences that contain at least one hit for the specific family hmm. 
We excluded those protein sequences containing undefined or ambiguous residues (X,B,Z,J).

\paragraph{Haloacid Dehalogenase globular family dataset (DEH$_f$).}
A globular family was retrieved from the SFLD site (http://sfld.rbvi.ucsf.edu/django/superfamily/3/) 
from which we selected the Haloacid Dehalogenase superfamily.
It was reduced to a 90\% identity for non redundancy with cd-hit. Once reduced, the total number of residues in that family was 24,031,515 which was a shorter amount of residues when compared to ANK$_f$ and WD$_f$ datasets. In order to be fair with all the datasets and avoid a bias due to random matches product of the dataset size, we reduced the ANK$_f$ and WD$_f$ to have an equivalent size to the DEH$_f$ ($\sim$24K residues). 

\paragraph{Heterogeneous dataset (HET$_f$).} We constructed a random  dataset 
 by  taking a sample of proteins from Uniref0.9 in such a way that the total size in number 
of residues was equivalent to the other datasets and the selected proteins do not belong 
to any of the above mentioned families. 

\paragraph{Scrambled heterogeneous dataset (HET$_f$ scrambled).} 
We constructed a new dataset,  scrambling the amino acids of each sequence from HET$_f$ dataset.

\subsection*{Protein test groups dataset} 
\paragraph{ANK test group dataset (ANK$_t$).} We retrieved all the non redundant structures, 
74 in total, from RepeatsDB \cite{DiDomenico}.

\paragraph{WD40 test group dataset (WD$_t$).} 50 Structures corresponding to members
 of the WD40 Protein Family, not included in the WD$_f$ set, 
were randomly selected in to conform this group.

\paragraph{Haloacid Dehalogenase test group dataset (DEH$_t$).} 50 Structures corresponding to 
members of the DEH Protein Family were randomly selected to conform this group. 
These structures were selected from the SFLD site, from those proteins that 
were not included when building the DEH$_f$ set. 

\paragraph{Globular Non Family test group dataset (HET$_t$).} 50 Structures corresponding 
to a set of unrelated globular proteins was used to conform this group \cite{Myers}.

\subsection*{Repeat finding algorithm}
The algorithm  {\em findpat}  \cite{Becher}, is the most current efficient algorithm to 
find exact repeats (it  particularly well suited for  very large inputs).
The algorithm requires a parameter $ml$ for the minimum length of a MR 
to be reported, it can be any value greater than or equal to $1$. 
For the special case  of  $ml$ equal to $0$ findpat returns all possible blocks of the given sequence. 
To avoid the use of multiple different special symbols  $\$_i$, for as many 
$i$ as needed, we modified the program to have an unique special symbol $\$$ 
as a symbol that can not be part of  MRs.
The algorithm {\em findpat} runs in time $O(n \log n)$, where $n$ 
is the length of the whole input (target sequence or sequences for the family of proteins).

\newpage
\section*{Supporting Information}

\setcounter{figure}{0}
\begin{figure}[ht!]
	\includegraphics[width=0.70\linewidth]{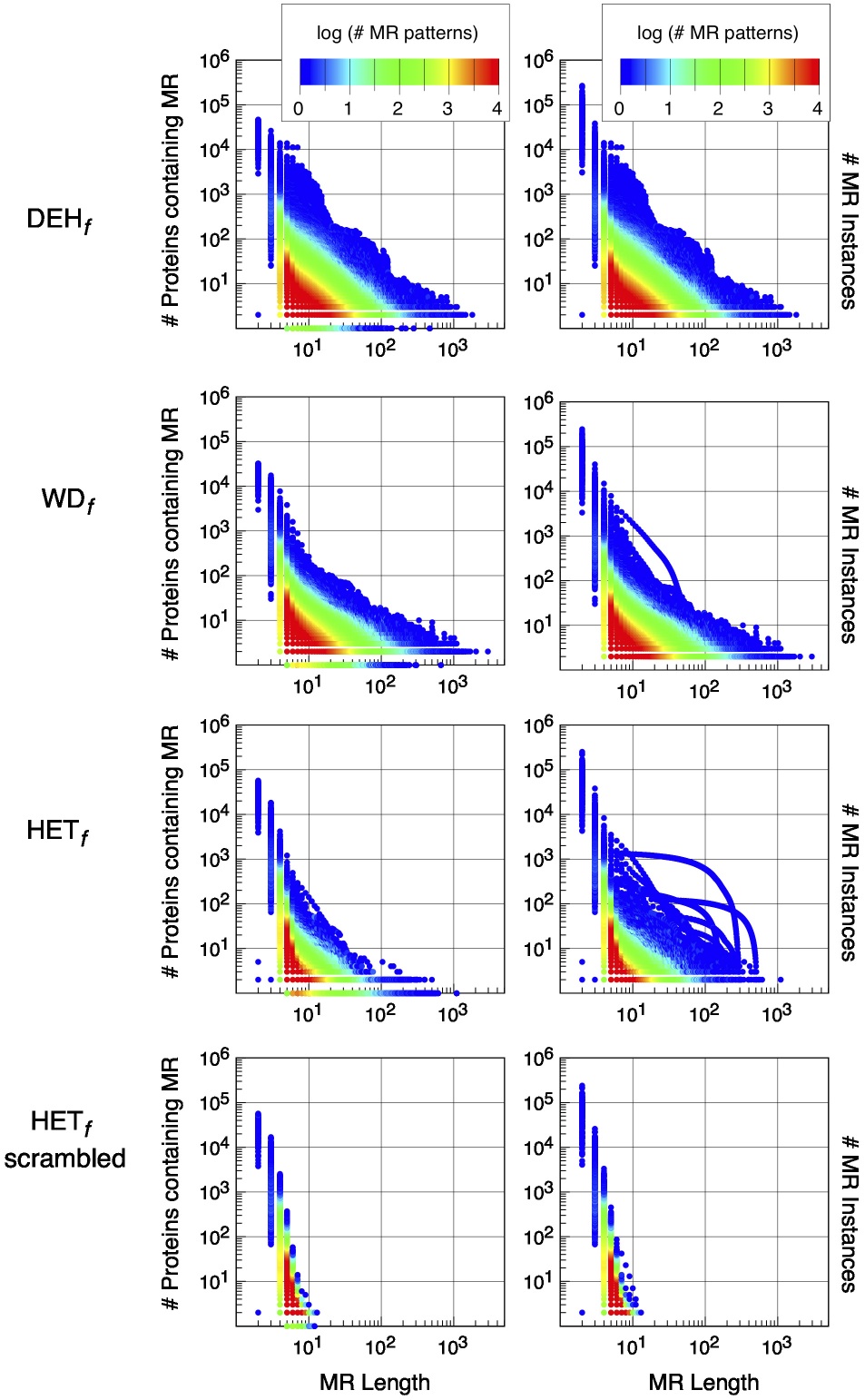}
	\caption*{{\bf Figure S1. Maximal repeats (MR) distribution within the proteins members of the family.}
	The sequences of distinct families were used to calculate the MR set. The distribution of the millions of MR found on the each set is shown according to the length of the pattern. On the left column, the number of different proteins that contain the MR pattern. On the right column, the number of time each MR pattern is present in the whole dataset. The colorscale denotes the number of different MR patterns that occur at a particular coordinate.}
\end{figure}

\clearpage
\newpage

\scriptsize
\begin{longtable}{|l|c|c|c|c|c|c|c|c|c|c|c|}
\caption*{{\bf Table S1. Coverage of trial sequences of the ANK$_{t}$ group.} The values for the $coverage(s,\M(s,i))$ function for each sequence, $s$, name by the UniProt identifier was calculated for different minimum length of the maximal repeats (MR) from $i=0\dots10$. Values truncated to two decimal places.}
\label{S1_Table}
\\
\hline
\multicolumn{ 1}{|c|}{\textbf{Uniprot ID}} & \multicolumn{ 1}{c|}{\textbf{$i=0$}}       & \multicolumn{ 1}{c|}{\textbf{$i=1$}}       & \multicolumn{ 1}{c|}{\textbf{$i=2$}}       & \multicolumn{ 1}{c|}{\textbf{$i=3$}}       & \multicolumn{ 1}{c|}{\textbf{$i=4$}}       & \multicolumn{ 1}{c|}{\textbf{$i=5$}}       & \multicolumn{ 1}{c|}{\textbf{$i=6$}}       & \multicolumn{ 1}{c|}{\textbf{$i=7$}}       & \multicolumn{ 1}{c|}{\textbf{$i=8$}}       & \multicolumn{ 1}{c|}{\textbf{$i=9$}}       & \multicolumn{ 1}{c|}{\textbf{$i=10$}}      \\
\multicolumn{ 1}{|c|}{\textbf{$(s)$}}      &                                            &                                            &                                            &                                            &                                            &                                            &                                            &                                            &                                            &                                            &                                            \\
\hline
DARPIN-1D5        &1.00&0.98&0.79&0.43&0.43&0.43&0.43&0.38&0.27&0.27&0.27\\ \hline
DARPIN-20         &1.00&0.99&0.77&0.40&0.40&0.40&0.32&0.27&0.16&0.16&0.16\\ \hline
DARPIN-3ANK       &1.00&0.98&0.98&0.98&0.98&0.98&0.98&0.98&0.98&0.98&0.98\\ \hline
DARPIN-3CA1A2N    &1.00&0.97&0.71&0.27&0.18&0.00&0.00&0.00&0.00&0.00&0.00\\ \hline
DARPIN-3CA1A2N-OH &1.00&0.97&0.75&0.32&0.23&0.10&0.00&0.00&0.00&0.00&0.00\\ \hline
DARPIN-3H10       &1.00&0.99&0.80&0.67&0.60&0.56&0.56&0.52&0.36&0.36&0.36\\ \hline
DARPIN-4ANK       &1.00&0.99&0.99&0.99&0.99&0.99&0.99&0.99&0.99&0.99&0.99\\ \hline
DARPIN-AR-3A      &1.00&1.00&0.78&0.49&0.49&0.49&0.40&0.32&0.23&0.23&0.23\\ \hline
DARPIN-AR-F8      &1.00&0.99&0.80&0.55&0.50&0.50&0.50&0.43&0.29&0.29&0.29\\ \hline
DARPIN-E3-19      &1.00&0.99&0.82&0.58&0.55&0.55&0.55&0.55&0.41&0.41&0.30\\ \hline
DARPIN-E3-5       &1.00&0.99&0.85&0.60&0.56&0.51&0.51&0.51&0.29&0.29&0.29\\ \hline
DARPIN-H10-2-G3   &1.00&0.99&0.80&0.37&0.32&0.32&0.16&0.11&0.00&0.00&0.00\\ \hline
DARPIN-NI1C-Mut4  &1.00&0.95&0.72&0.20&0.08&0.00&0.00&0.00&0.00&0.00&0.00\\ \hline
DARPIN-NI3C       &1.00&0.98&0.88&0.70&0.68&0.68&0.68&0.68&0.68&0.68&0.68\\ \hline
DARPIN-NI3C-Mut5  &1.00&0.99&0.89&0.70&0.66&0.66&0.66&0.66&0.66&0.66&0.66\\ \hline
DARPIN-OFF7       &1.00&1.00&0.79&0.50&0.50&0.50&0.50&0.49&0.36&0.26&0.26\\ \hline
E9ADW8            &1.00&1.00&0.88&0.26&0.06&0.00&0.00&0.00&0.00&0.00&0.00\\ \hline
NRC               &1.00&0.99&0.95&0.89&0.77&0.73&0.68&0.59&0.59&0.45&0.35\\ \hline
O14593            &1.00&1.00&0.82&0.25&0.00&0.00&0.00&0.00&0.00&0.00&0.00\\ \hline
O22265            &1.00&1.00&0.85&0.25&0.09&0.04&0.01&0.00&0.00&0.00&0.00\\ \hline
O35433            &1.00&1.00&0.97&0.37&0.03&0.00&0.00&0.00&0.00&0.00&0.00\\ \hline
O75832            &1.00&0.99&0.81&0.16&0.03&0.00&0.00&0.00&0.00&0.00&0.00\\ \hline
OR264             &1.00&0.98&0.89&0.81&0.76&0.76&0.70&0.63&0.63&0.63&0.57\\ \hline
OR266             &1.00&0.98&0.89&0.77&0.71&0.71&0.60&0.57&0.57&0.57&0.57\\ \hline
P07207            &1.00&1.00&0.99&0.80&0.36&0.19&0.13&0.08&0.02&0.01&0.01\\ \hline
P09959            &1.00&1.00&0.97&0.50&0.08&0.01&0.00&0.00&0.00&0.00&0.00\\ \hline
P14585            &1.00&1.00&0.99&0.58&0.11&0.02&0.00&0.00&0.00&0.00&0.00\\ \hline
P16157            &1.00&1.00&0.99&0.75&0.28&0.12&0.08&0.04&0.02&0.00&0.00\\ \hline
P20749            &1.00&1.00&0.91&0.48&0.13&0.02&0.00&0.00&0.00&0.00&0.00\\ \hline
P25963            &1.00&1.00&0.84&0.32&0.16&0.08&0.03&0.00&0.00&0.00&0.00\\ \hline
P42771            &1.00&0.98&0.82&0.21&0.00&0.00&0.00&0.00&0.00&0.00&0.00\\ \hline
P42773            &1.00&0.98&0.77&0.06&0.00&0.00&0.00&0.00&0.00&0.00&0.00\\ \hline
P46531            &1.00&1.00&0.99&0.82&0.36&0.21&0.08&0.03&0.01&0.00&0.00\\ \hline
P46683            &1.00&0.99&0.80&0.07&0.04&0.00&0.00&0.00&0.00&0.00&0.00\\ \hline
P50086            &1.00&1.00&0.83&0.20&0.07&0.00&0.00&0.00&0.00&0.00&0.00\\ \hline
P55271            &1.00&0.98&0.70&0.16&0.00&0.00&0.00&0.00&0.00&0.00&0.00\\ \hline
P55273            &1.00&1.00&0.83&0.23&0.00&0.00&0.00&0.00&0.00&0.00&0.00\\ \hline
P58546            &1.00&0.99&0.49&0.15&0.06&0.00&0.00&0.00&0.00&0.00&0.00\\ \hline
P62774            &1.00&0.99&0.51&0.15&0.06&0.00&0.00&0.00&0.00&0.00&0.00\\ \hline
P62775            &1.00&0.99&0.51&0.15&0.06&0.00&0.00&0.00&0.00&0.00&0.00\\ \hline
Q00420            &1.00&1.00&0.88&0.28&0.06&0.00&0.00&0.00&0.00&0.00&0.00\\ \hline
Q01705            &1.00&1.00&0.99&0.82&0.37&0.21&0.09&0.04&0.01&0.00&0.00\\ \hline
Q05823            &1.00&1.00&0.97&0.51&0.08&0.03&0.00&0.00&0.00&0.00&0.00\\ \hline
Q13418            &1.00&1.00&0.92&0.18&0.01&0.00&0.00&0.00&0.00&0.00&0.00\\ \hline
Q13625            &1.00&1.00&0.98&0.58&0.11&0.01&0.00&0.00&0.00&0.00&0.00\\ \hline
Q15027            &1.00&1.00&0.96&0.43&0.07&0.00&0.00&0.00&0.00&0.00&0.00\\ \hline
Q5ZSV0            &1.00&1.00&0.83&0.41&0.25&0.21&0.19&0.17&0.17&0.13&0.13\\ \hline
Q5ZXN6            &1.00&1.00&0.98&0.50&0.08&0.01&0.00&0.00&0.00&0.00&0.00\\ \hline
Q60773            &1.00&0.99&0.71&0.12&0.04&0.00&0.00&0.00&0.00&0.00&0.00\\ \hline
Q60778            &1.00&1.00&0.90&0.36&0.18&0.09&0.05&0.00&0.00&0.00&0.00\\ \hline
Q63ZY3            &1.00&1.00&0.97&0.51&0.09&0.02&0.01&0.01&0.00&0.00&0.00\\ \hline
Q6IV60            &1.00&1.00&0.83&0.22&0.02&0.00&0.00&0.00&0.00&0.00&0.00\\ \hline
Q6PFX9            &1.00&1.00&0.98&0.63&0.33&0.19&0.15&0.13&0.08&0.07&0.05\\ \hline
Q7SIG6            &1.00&1.00&0.98&0.49&0.07&0.02&0.00&0.00&0.00&0.00&0.00\\ \hline
Q838Q8            &1.00&1.00&0.83&0.31&0.03&0.00&0.00&0.00&0.00&0.00&0.00\\ \hline
Q8IUH5            &1.00&1.00&0.94&0.32&0.01&0.00&0.00&0.00&0.00&0.00&0.00\\ \hline
Q8TDY4            &1.00&1.00&0.97&0.47&0.04&0.00&0.00&0.00&0.00&0.00&0.00\\ \hline
Q8WUF5            &1.00&1.00&0.96&0.56&0.21&0.05&0.02&0.01&0.01&0.01&0.00\\ \hline
Q90623            &1.00&1.00&0.98&0.62&0.25&0.08&0.03&0.00&0.00&0.00&0.00\\ \hline
Q91WD2            &1.00&1.00&0.97&0.30&0.02&0.01&0.01&0.00&0.00&0.00&0.00\\ \hline
Q92882            &1.00&1.00&0.69&0.13&0.00&0.00&0.00&0.00&0.00&0.00&0.00\\ \hline
Q96DX5            &1.00&1.00&0.87&0.29&0.05&0.00&0.00&0.00&0.00&0.00&0.00\\ \hline
Q96NW4            &1.00&1.00&0.98&0.46&0.10&0.06&0.03&0.02&0.00&0.00&0.00\\ \hline
Q978J0            &1.00&0.98&0.84&0.26&0.04&0.00&0.00&0.00&0.00&0.00&0.00\\ \hline
Q99728            &1.00&1.00&0.96&0.35&0.05&0.02&0.00&0.00&0.00&0.00&0.00\\ \hline
Q9DFS3            &1.00&1.00&0.98&0.43&0.04&0.00&0.00&0.00&0.00&0.00&0.00\\ \hline
Q9H2K2            &1.00&1.00&0.98&0.60&0.32&0.23&0.21&0.17&0.15&0.10&0.07\\ \hline
Q9H9B1            &1.00&1.00&0.99&0.53&0.09&0.00&0.00&0.00&0.00&0.00&0.00\\ \hline
Q9H9E1            &1.00&1.00&0.85&0.19&0.10&0.03&0.00&0.00&0.00&0.00&0.00\\ \hline
Q9HBA0            &1.00&1.00&0.97&0.38&0.03&0.01&0.00&0.00&0.00&0.00&0.00\\ \hline
Q9WUD2            &1.00&1.00&0.96&0.43&0.07&0.00&0.00&0.00&0.00&0.00&0.00\\ \hline
Q9Y5S1            &1.00&1.00&0.96&0.42&0.06&0.00&0.00&0.00&0.00&0.00&0.00\\ \hline
Q9Z2X2            &1.00&1.00&0.77&0.18&0.03&0.00&0.00&0.00&0.00&0.00&0.00\\ \hline
\end{longtable}

\clearpage
\newpage

\scriptsize
\begin{longtable}{|l|c|c|c|c|c|c|c|c|c|c|c|}
\caption*{{\bf Table S2. Coverage of trial sequences of the DEH$_{t}$ group.} The values for the $coverage(s,\M(s,i))$ function for each sequence, $s$, name by the UniProt identifier was calculated for different minimum length of the maximal repeats (MR) from $i=0\dots10$. Values truncated to two decimal places.}
\label{S2_Table}
\\
\hline
\multicolumn{ 1}{|c|}{\textbf{Uniprot ID}} & \multicolumn{ 1}{c|}{\textbf{$i=0$}}       & \multicolumn{ 1}{c|}{\textbf{$i=1$}}       & \multicolumn{ 1}{c|}{\textbf{$i=2$}}       & \multicolumn{ 1}{c|}{\textbf{$i=3$}}       & \multicolumn{ 1}{c|}{\textbf{$i=4$}}       & \multicolumn{ 1}{c|}{\textbf{$i=5$}}       & \multicolumn{ 1}{c|}{\textbf{$i=6$}}       & \multicolumn{ 1}{c|}{\textbf{$i=7$}}       & \multicolumn{ 1}{c|}{\textbf{$i=8$}}       & \multicolumn{ 1}{c|}{\textbf{$i=9$}}       & \multicolumn{ 1}{c|}{\textbf{$i=10$}}      \\
\multicolumn{ 1}{|c|}{\textbf{$(s)$}}      &                                            &                                            &                                            &                                            &                                            &                                            &                                            &                                            &                                            &                                            &                                            \\
\hline
A5LNI9&1.00&1.00&0.83&0.23&0.07&0.03&0.00&0.00&0.00&0.00&0.00\\ \hline
B0A4S9&1.00&0.98&0.77&0.17&0.00&0.00&0.00&0.00&0.00&0.00&0.00\\ \hline
B2Z3V8&1.00&0.99&0.69&0.15&0.00&0.00&0.00&0.00&0.00&0.00&0.00\\ \hline
B4A833&1.00&1.00&0.79&0.16&0.00&0.00&0.00&0.00&0.00&0.00&0.00\\ \hline
B4ABS1&1.00&1.00&0.81&0.21&0.03&0.00&0.00&0.00&0.00&0.00&0.00\\ \hline
C2JJ26&1.00&1.00&0.88&0.21&0.09&0.00&0.00&0.00&0.00&0.00&0.00\\ \hline
C6IG92&1.00&0.99&0.56&0.00&0.00&0.00&0.00&0.00&0.00&0.00&0.00\\ \hline
C7RF86&1.00&1.00&0.88&0.27&0.04&0.02&0.00&0.00&0.00&0.00&0.00\\ \hline
D4FVT8&1.00&1.00&0.69&0.00&0.00&0.00&0.00&0.00&0.00&0.00&0.00\\ \hline
D4V4M5&1.00&0.99&0.71&0.20&0.03&0.00&0.00&0.00&0.00&0.00&0.00\\ \hline
D7IJ01&1.00&0.99&0.80&0.17&0.03&0.00&0.00&0.00&0.00&0.00&0.00\\ \hline
D8FP15&1.00&0.99&0.77&0.11&0.00&0.00&0.00&0.00&0.00&0.00&0.00\\ \hline
G9RLJ0&1.00&1.00&0.78&0.09&0.00&0.00&0.00&0.00&0.00&0.00&0.00\\ \hline
K2T267&1.00&0.99&0.72&0.13&0.00&0.00&0.00&0.00&0.00&0.00&0.00\\ \hline
M2DKS0&1.00&0.99&0.89&0.22&0.04&0.00&0.00&0.00&0.00&0.00&0.00\\ \hline
M4SEQ9&1.00&1.00&0.96&0.47&0.05&0.00&0.00&0.00&0.00&0.00&0.00\\ \hline
O08575&1.00&1.00&0.91&0.31&0.06&0.00&0.00&0.00&0.00&0.00&0.00\\ \hline
O15305&1.00&0.99&0.76&0.20&0.03&0.00&0.00&0.00&0.00&0.00&0.00\\ \hline
O29777&1.00&1.00&0.97&0.55&0.09&0.02&0.01&0.00&0.00&0.00&0.00\\ \hline
O32125&1.00&0.99&0.79&0.09&0.02&0.02&0.00&0.00&0.00&0.00&0.00\\ \hline
O32220&1.00&1.00&0.98&0.51&0.10&0.01&0.01&0.01&0.01&0.00&0.00\\ \hline
O59346&1.00&1.00&0.78&0.12&0.00&0.00&0.00&0.00&0.00&0.00&0.00\\ \hline
O67920&1.00&1.00&0.74&0.23&0.00&0.00&0.00&0.00&0.00&0.00&0.00\\ \hline
P05023&1.00&1.00&0.97&0.30&0.08&0.01&0.00&0.00&0.00&0.00&0.00\\ \hline
P06685&1.00&1.00&0.99&0.43&0.05&0.00&0.00&0.00&0.00&0.00&0.00\\ \hline
P0AE22&1.00&0.99&0.72&0.12&0.00&0.00&0.00&0.00&0.00&0.00&0.00\\ \hline
P20649&1.00&1.00&0.98&0.42&0.03&0.01&0.01&0.00&0.00&0.00&0.00\\ \hline
P35670&1.00&0.99&0.71&0.04&0.00&0.00&0.00&0.00&0.00&0.00&0.00\\ \hline
P78330&1.00&0.99&0.74&0.11&0.00&0.00&0.00&0.00&0.00&0.00&0.00\\ \hline
P94592&1.00&1.00&0.83&0.08&0.00&0.00&0.00&0.00&0.00&0.00&0.00\\ \hline
Q04656&1.00&0.96&0.52&0.13&0.06&0.06&0.00&0.00&0.00&0.00&0.00\\ \hline
Q11S56&1.00&0.99&0.76&0.14&0.02&0.00&0.00&0.00&0.00&0.00&0.00\\ \hline
Q2T109&1.00&0.98&0.67&0.02&0.00&0.00&0.00&0.00&0.00&0.00&0.00\\ \hline
Q3UGR5&1.00&1.00&0.83&0.24&0.00&0.00&0.00&0.00&0.00&0.00&0.00\\ \hline
Q5EBQ9&1.00&0.99&0.73&0.05&0.00&0.00&0.00&0.00&0.00&0.00&0.00\\ \hline
Q5SJQ3&1.00&0.99&0.90&0.39&0.18&0.03&0.00&0.00&0.00&0.00&0.00\\ \hline
Q60048&1.00&1.00&0.97&0.44&0.02&0.00&0.00&0.00&0.00&0.00&0.00\\ \hline
Q7ADF8&1.00&1.00&0.75&0.12&0.07&0.00&0.00&0.00&0.00&0.00&0.00\\ \hline
Q7WG29&1.00&1.00&0.62&0.23&0.00&0.00&0.00&0.00&0.00&0.00&0.00\\ \hline
Q8K7R3&1.00&0.99&0.85&0.24&0.03&0.00&0.00&0.00&0.00&0.00&0.00\\ \hline
Q8L1N9&1.00&0.99&0.74&0.25&0.03&0.00&0.00&0.00&0.00&0.00&0.00\\ \hline
Q8TBE9&1.00&1.00&0.80&0.13&0.05&0.02&0.00&0.00&0.00&0.00&0.00\\ \hline
Q96X90&1.00&1.00&0.85&0.20&0.00&0.00&0.00&0.00&0.00&0.00&0.00\\ \hline
Q96XE7&1.00&1.00&0.83&0.13&0.03&0.00&0.00&0.00&0.00&0.00&0.00\\ \hline
Q98I56&1.00&0.99&0.75&0.17&0.03&0.00&0.00&0.00&0.00&0.00&0.00\\ \hline
Q9D020&1.00&0.99&0.85&0.19&0.00&0.00&0.00&0.00&0.00&0.00&0.00\\ \hline
Q9JLV6&1.00&1.00&0.93&0.32&0.01&0.00&0.00&0.00&0.00&0.00&0.00\\ \hline
Q9X0Y1&1.00&1.00&0.85&0.25&0.12&0.04&0.00&0.00&0.00&0.00&0.00\\ \hline
T0QBN5&1.00&0.99&0.79&0.16&0.03&0.00&0.00&0.00&0.00&0.00&0.00\\ \hline
U8H1V1&1.00&1.00&0.73&0.15&0.00&0.00&0.00&0.00&0.00&0.00&0.00\\ \hline
\end{longtable}
\clearpage
\newpage

\scriptsize
\begin{longtable}{|l|c|c|c|c|c|c|c|c|c|c|c|}
\caption*{{\bf Table S3. Coverage of trial sequences of the WD$_{t}$ group.} The values for the $coverage(s,\M(s,i))$ function for each sequence, $s$, name by the UniProt identifier was calculated for different minimum length of the maximal repeats (MR) from $i=0\dots10$. Values truncated to two decimal places.}
\label{S3_Table}
\\
\hline
\multicolumn{ 1}{|c|}{\textbf{Uniprot ID}} & \multicolumn{ 1}{c|}{\textbf{$i=0$}}       & \multicolumn{ 1}{c|}{\textbf{$i=1$}}       & \multicolumn{ 1}{c|}{\textbf{$i=2$}}       & \multicolumn{ 1}{c|}{\textbf{$i=3$}}       & \multicolumn{ 1}{c|}{\textbf{$i=4$}}       & \multicolumn{ 1}{c|}{\textbf{$i=5$}}       & \multicolumn{ 1}{c|}{\textbf{$i=6$}}       & \multicolumn{ 1}{c|}{\textbf{$i=7$}}       & \multicolumn{ 1}{c|}{\textbf{$i=8$}}       & \multicolumn{ 1}{c|}{\textbf{$i=9$}}       & \multicolumn{ 1}{c|}{\textbf{$i=10$}}      \\
\multicolumn{ 1}{|c|}{\textbf{$(s)$}}      &                                            &                                            &                                            &                                            &                                            &                                            &                                            &                                            &                                            &                                            &                                            \\
\hline
A6ZU46&1.00&1.00&0.98&0.44&0.04&0.01&0.00&0.00&0.00&0.00&0.00\\ \hline
O14727&1.00&1.00&0.99&0.52&0.08&0.03&0.00&0.00&0.00&0.00&0.00\\ \hline
O24456&1.00&1.00&0.85&0.38&0.11&0.08&0.04&0.04&0.00&0.00&0.00\\ \hline
O75530&1.00&1.00&0.90&0.10&0.00&0.00&0.00&0.00&0.00&0.00&0.00\\ \hline
O76071&1.00&0.99&0.85&0.36&0.17&0.12&0.07&0.04&0.00&0.00&0.00\\ \hline
O88879&1.00&1.00&0.99&0.53&0.09&0.00&0.00&0.00&0.00&0.00&0.00\\ \hline
O89053&1.00&1.00&0.90&0.27&0.08&0.02&0.00&0.00&0.00&0.00&0.00\\ \hline
P07834&1.00&1.00&0.97&0.49&0.15&0.04&0.01&0.00&0.00&0.00&0.00\\ \hline
P16649&1.00&1.00&0.97&0.52&0.22&0.07&0.04&0.01&0.01&0.00&0.00\\ \hline
P26449&1.00&1.00&0.87&0.17&0.02&0.00&0.00&0.00&0.00&0.00&0.00\\ \hline
P36037&1.00&1.00&0.96&0.39&0.03&0.00&0.00&0.00&0.00&0.00&0.00\\ \hline
P38011&1.00&1.00&0.79&0.39&0.14&0.03&0.00&0.00&0.00&0.00&0.00\\ \hline
P38262&1.00&1.00&0.93&0.36&0.04&0.00&0.00&0.00&0.00&0.00&0.00\\ \hline
P38968&1.00&1.00&0.98&0.61&0.17&0.06&0.02&0.01&0.00&0.00&0.00\\ \hline
P40217&1.00&1.00&0.89&0.21&0.06&0.00&0.00&0.00&0.00&0.00&0.00\\ \hline
P46680&1.00&1.00&0.94&0.36&0.06&0.00&0.00&0.00&0.00&0.00&0.00\\ \hline
P53011&1.00&1.00&0.87&0.18&0.00&0.00&0.00&0.00&0.00&0.00&0.00\\ \hline
P53196&1.00&1.00&0.86&0.16&0.03&0.00&0.00&0.00&0.00&0.00&0.00\\ \hline
P54311&1.00&1.00&0.87&0.28&0.08&0.02&0.00&0.00&0.00&0.00&0.00\\ \hline
P55735&1.00&1.00&0.85&0.26&0.13&0.00&0.00&0.00&0.00&0.00&0.00\\ \hline
P61964&1.00&0.99&0.87&0.43&0.23&0.14&0.10&0.08&0.00&0.00&0.00\\ \hline
P61965&1.00&0.99&0.87&0.43&0.23&0.14&0.10&0.08&0.00&0.00&0.00\\ \hline
P62871&1.00&1.00&0.87&0.28&0.08&0.02&0.00&0.00&0.00&0.00&0.00\\ \hline
P62881&1.00&1.00&0.87&0.30&0.02&0.00&0.00&0.00&0.00&0.00&0.00\\ \hline
P63005&1.00&1.00&0.89&0.34&0.11&0.04&0.04&0.04&0.04&0.04&0.00\\ \hline
P63244&1.00&1.00&0.87&0.31&0.08&0.03&0.03&0.00&0.00&0.00&0.00\\ \hline
P78406&1.00&1.00&0.85&0.14&0.00&0.00&0.00&0.00&0.00&0.00&0.00\\ \hline
P78774&1.00&1.00&0.91&0.22&0.05&0.00&0.00&0.00&0.00&0.00&0.00\\ \hline
P78972&1.00&1.00&0.95&0.31&0.03&0.00&0.00&0.00&0.00&0.00&0.00\\ \hline
Q02793&1.00&1.00&0.87&0.27&0.02&0.00&0.00&0.00&0.00&0.00&0.00\\ \hline
Q03774&1.00&1.00&0.91&0.30&0.01&0.00&0.00&0.00&0.00&0.00&0.00\\ \hline
Q04491&1.00&0.99&0.87&0.27&0.02&0.00&0.00&0.00&0.00&0.00&0.00\\ \hline
Q04724&1.00&1.00&0.96&0.40&0.07&0.00&0.00&0.00&0.00&0.00&0.00\\ \hline
Q05583&1.00&1.00&0.86&0.24&0.05&0.03&0.00&0.00&0.00&0.00&0.00\\ \hline
Q09028&1.00&1.00&0.90&0.29&0.08&0.02&0.00&0.00&0.00&0.00&0.00\\ \hline
Q11176&1.00&1.00&0.97&0.40&0.07&0.00&0.00&0.00&0.00&0.00&0.00\\ \hline
Q13216&1.00&1.00&0.88&0.26&0.04&0.00&0.00&0.00&0.00&0.00&0.00\\ \hline
Q16576&1.00&1.00&0.89&0.23&0.01&0.00&0.00&0.00&0.00&0.00&0.00\\ \hline
Q24572&1.00&1.00&0.90&0.21&0.06&0.02&0.00&0.00&0.00&0.00&0.00\\ \hline
Q24D42&1.00&1.00&0.86&0.11&0.02&0.02&0.00&0.00&0.00&0.00&0.00\\ \hline
Q2YDS1&1.00&1.00&0.92&0.33&0.01&0.01&0.00&0.00&0.00&0.00&0.00\\ \hline
Q58CQ2&1.00&1.00&0.86&0.19&0.02&0.00&0.00&0.00&0.00&0.00&0.00\\ \hline
Q6CN23&1.00&1.00&0.84&0.18&0.04&0.00&0.00&0.00&0.00&0.00&0.00\\ \hline
Q8LNY6&1.00&1.00&0.86&0.28&0.00&0.00&0.00&0.00&0.00&0.00&0.00\\ \hline
Q921E6&1.00&1.00&0.90&0.10&0.00&0.00&0.00&0.00&0.00&0.00&0.00\\ \hline
Q92466&1.00&1.00&0.89&0.13&0.03&0.00&0.00&0.00&0.00&0.00&0.00\\ \hline
Q969H0&1.00&1.00&0.95&0.49&0.13&0.09&0.00&0.00&0.00&0.00&0.00\\ \hline
Q96MX6&1.00&1.00&0.84&0.19&0.02&0.00&0.00&0.00&0.00&0.00&0.00\\ \hline
Q9GZS3&1.00&1.00&0.90&0.29&0.09&0.03&0.03&0.00&0.00&0.00&0.00\\ \hline
Q9Y297&1.00&1.00&0.93&0.35&0.10&0.01&0.00&0.00&0.00&0.00&0.00\\ \hline
\end{longtable}
\clearpage
\newpage

\scriptsize
\begin{longtable}{|l|l|c|c|c|cHH|c|}
\caption*{{\bf Table S4. Values of \em{familiarity} \bf function for several proteins and families.} Values of $familiarity(s,t)$ function for $s$ proteins from ANK$_{t}$, DEH$_{t}$, HET$_{t}$ and WD$_{t}$ test group dataset and $t$=$s$, ANK$_{f}$, DEH$_{f}$, HET$_{f}$ and WD$_{f}$.}
\label{S4_Table}
\\
\hline
\multicolumn{ 1}{|c|}{\textbf{Uniprot ID}} & \multicolumn{ 1}{c|}{\textbf{Test}}   & \multicolumn{ 1}{c|}{\textbf{familiarity}} & \multicolumn{ 1}{c|}{\textbf{familiarity}} & \multicolumn{ 1}{c|}{\textbf{familiarity}} & \multicolumn{ 1}{|c}{\textbf{familiarity}}   & \multicolumn{ 1}{H}{\textbf{familiarity}}            & \multicolumn{ 1}{H|}{\textbf{familiarity}}                & \multicolumn{ 1}{c|}{\textbf{familiarity}} \\
\multicolumn{ 1}{|c|}{\textbf{$(s)$}}      & \multicolumn{ 1}{|c|}{\textbf{Group}} & \multicolumn{ 1}{c|}{\textbf{(s,s)}}       & \multicolumn{ 1}{c|}{\textbf{(s,ANK$_f$)}} & \multicolumn{ 1}{c|}{\textbf{(s,DEH$_f$)}} & \multicolumn{ 1}{|c}{\textbf{(s,HET$_f$)}}   & \multicolumn{ 1}{H}{\textbf{(s,MixedScrambled$_f$)}} & \multicolumn{ 1}{H|}{\textbf{(s,MixedRandomUniform$_f$)}} & \multicolumn{ 1}{c|}{\textbf{(s,WD$_f$)}}  \\
\hline
DARPIN-1D5       &ANK$_t$&5.07600&9.28800&6.09600&6.38400&6.25600&5.75200&6.44000\\ \hline
DARPIN-20        &ANK$_t$&4.47984&9.25806&6.13306&6.43952&6.39919&5.84274&6.22984\\ \hline
DARPIN-3ANK      &ANK$_t$&9.90110&9.89560&6.34066&6.42857&6.58242&5.78022&6.98901\\ \hline
DARPIN-3CA1A2N   &ANK$_t$&2.65909&9.13068&6.25568&6.52841&6.48295&5.72159&6.58523\\ \hline
DARPIN-3CA1A2N-OH&ANK$_t$&2.90625&8.82292&6.26562&6.57813&6.43229&5.79688&6.69271\\ \hline
DARPIN-3H10      &ANK$_t$&6.14615&9.39231&6.11923&6.46538&6.45000&5.68846&6.49615\\ \hline
DARPIN-4ANK      &ANK$_t$&9.92857&9.92857&6.34524&6.45635&6.57540&5.84524&6.98810\\ \hline
DARPIN-AR-3A     &ANK$_t$&5.08065&8.90645&5.93548&6.36774&6.38065&5.71613&6.33548\\ \hline
DARPIN-AR-F8     &ANK$_t$&5.55063&9.34494&5.95886&6.40190&6.38291&5.69937&6.39557\\ \hline
DARPIN-E3-19     &ANK$_t$&6.11290&9.01935&5.89677&6.45806&6.45806&5.72903&6.50323\\ \hline
DARPIN-E3-5      &ANK$_t$&5.80818&9.15409&6.23270&6.33962&6.32075&5.66038&6.45283\\ \hline
DARPIN-H10-2-G3  &ANK$_t$&3.59274&8.77419&6.25403&6.47984&6.39113&5.70565&6.25403\\ \hline
DARPIN-NI1C-Mut4 &ANK$_t$&2.50000&9.20652&6.26630&6.64674&6.45109&5.64674&6.25543\\ \hline
DARPIN-NI3C      &ANK$_t$&7.51623&9.61039&6.32143&6.67857&6.43182&5.66558&6.41234\\ \hline
DARPIN-NI3C-Mut5 &ANK$_t$&7.44268&9.23567&6.40764&6.66242&6.40764&5.70064&6.40764\\ \hline
DARPIN-OFF7      &ANK$_t$&5.58599&9.31847&6.05732&6.28662&6.41401&5.71338&6.45860\\ \hline
E9ADW8           &ANK$_t$&2.71111&8.21250&6.04861&6.18750&6.12361&5.74028&5.97917\\ \hline
NRC              &ANK$_t$&7.36087&8.73913&6.17391&6.11304&6.19130&5.56522&6.29565\\ \hline
O14593           &ANK$_t$&2.58462&9.23077&6.07885&6.37115&6.28654&5.82500&6.14038\\ \hline
O22265           &ANK$_t$&2.77614&7.49330&6.12601&6.36997&6.29759&5.78016&6.32976\\ \hline
O35433           &ANK$_t$&2.88425&8.90155&5.96957&6.17959&6.18675&5.72852&6.03640\\ \hline
O75832           &ANK$_t$&2.51106&9.96018&6.01549&6.24558&6.13938&5.70575&6.09513\\ \hline
OR264            &ANK$_t$&7.63018&8.97337&6.33728&6.47337&6.34320&5.89941&6.42012\\ \hline
OR266            &ANK$_t$&7.20414&8.98225&6.40237&6.49704&6.25444&5.82840&6.60947\\ \hline
P07207           &ANK$_t$&4.13559&8.88124&5.84221&6.17888&5.96152&5.78468&5.99649\\ \hline
P09959           &ANK$_t$&3.07659&8.63636&6.06102&6.19552&6.15567&5.71108&6.08593\\ \hline
P14585           &ANK$_t$&3.22638&6.70329&5.71099&5.98880&5.94262&5.77257&5.80616\\ \hline
P16157           &ANK$_t$&3.81366&9.88676&6.03615&6.21425&6.25784&5.72568&6.17916\\ \hline
P20749           &ANK$_t$&3.06388&9.61013&6.15969&6.37996&6.33811&5.75441&6.31608\\ \hline
P25963           &ANK$_t$&2.95110&9.57571&5.98423&6.14196&6.15773&5.69401&6.06309\\ \hline
P42771           &ANK$_t$&2.53526&9.59295&6.36859&6.56090&6.38782&5.68910&6.29167\\ \hline
P42773           &ANK$_t$&2.33036&8.84226&6.02679&6.31250&6.31250&5.54464&6.18750\\ \hline
P46531           &ANK$_t$&4.04090&9.95362&5.71820&6.17143&5.95969&5.81331&5.91781\\ \hline
P46683           &ANK$_t$&2.40750&8.77000&6.05250&6.23750&6.22750&5.78250&6.15750\\ \hline
P50086           &ANK$_t$&2.60965&8.67982&5.94079&6.05482&6.10307&5.74781&6.02412\\ \hline
P55271           &ANK$_t$&2.36154&9.32308&6.25769&6.49615&6.33462&5.68846&6.25000\\ \hline
P55273           &ANK$_t$&2.57229&9.14759&6.45482&6.50904&6.37651&5.73795&6.26807\\ \hline
P58546           &ANK$_t$&2.20763&9.82627&6.12288&6.43644&6.21610&5.91102&6.13136\\ \hline
P62774           &ANK$_t$&2.23305&9.82627&6.13136&6.43644&6.21610&5.91102&6.16525\\ \hline
P62775           &ANK$_t$&2.23305&9.82627&6.13136&6.43644&6.21610&5.91102&6.16525\\ \hline
Q00420           &ANK$_t$&2.73760&9.67885&6.09661&6.29504&6.32637&5.75196&6.18277\\ \hline
Q01705           &ANK$_t$&4.07171&9.23252&5.70170&6.12999&5.93441&5.79060&5.89648\\ \hline
Q05823           &ANK$_t$&3.10324&8.65385&6.07018&6.25371&6.12821&5.73954&6.07152\\ \hline
Q13418           &ANK$_t$&2.62168&9.74889&5.94580&6.05642&6.10730&5.70243&5.97456\\ \hline
Q13625           &ANK$_t$&3.19947&9.96144&6.00133&6.28324&6.19016&5.74512&6.09087\\ \hline
Q15027           &ANK$_t$&2.97432&9.92365&6.05068&6.32905&6.28581&5.69932&6.16824\\ \hline
Q5ZSV0           &ANK$_t$&3.94565&6.35462&6.03940&6.30842&6.14266&5.80842&5.98777\\ \hline
Q5ZXN6           &ANK$_t$&3.08588&6.34721&6.02529&6.26554&6.22234&5.72076&6.03793\\ \hline
Q60773           &ANK$_t$&2.38253&9.17470&6.34036&6.36446&6.26807&5.79217&6.31627\\ \hline
Q60778           &ANK$_t$&3.09331&8.96518&6.22841&6.40111&6.27577&5.75766&6.21170\\ \hline
Q63ZY3           &ANK$_t$&3.14160&9.39424&6.13043&6.36193&6.33020&5.66275&6.25264\\ \hline
Q6IV60           &ANK$_t$&2.59507&6.31514&6.05106&6.14965&6.23415&5.81866&5.97711\\ \hline
Q6PFX9           &ANK$_t$&4.12500&9.71477&6.24356&6.40038&6.28977&5.75038&6.24811\\ \hline
Q7SIG6           &ANK$_t$&3.06994&9.54593&5.99426&6.19468&6.21242&5.77192&6.08403\\ \hline
Q838Q8           &ANK$_t$&2.68905&8.26866&6.04975&6.24378&6.11443&5.78607&6.07960\\ \hline
Q8IUH5           &ANK$_t$&2.78956&9.77532&5.82674&5.98022&5.95174&5.70965&5.85047\\ \hline
Q8TDY4           &ANK$_t$&2.99612&9.68328&6.07752&6.23256&6.23367&5.75748&6.19712\\ \hline
Q8WUF5           &ANK$_t$&3.36353&9.60145&6.17935&6.43176&6.26993&5.71920&6.28140\\ \hline
Q90623           &ANK$_t$&3.49203&9.91833&6.22759&6.47062&6.30627&5.75548&6.36504\\ \hline
Q91WD2           &ANK$_t$&2.83700&9.22696&5.98900&6.20770&6.12380&5.74415&5.99312\\ \hline
Q92882           &ANK$_t$&2.32243&9.75000&6.12850&6.20327&6.24533&5.70327&5.97430\\ \hline
Q96DX5           &ANK$_t$&2.72109&9.26361&5.86565&6.12755&6.19218&5.74660&6.07313\\ \hline
Q96NW4           &ANK$_t$&3.17714&9.86667&5.94238&6.15095&6.16048&5.76810&6.07000\\ \hline
Q978J0           &ANK$_t$&2.65344&7.29101&6.13757&6.23810&6.18519&5.75132&6.05820\\ \hline
Q99728           &ANK$_t$&2.90669&9.96589&6.01030&6.12098&6.14801&5.75032&6.08623\\ \hline
Q9DFS3           &ANK$_t$&2.96596&9.47300&6.04049&6.18369&6.16843&5.75293&6.03228\\ \hline
Q9H2K2           &ANK$_t$&4.34991&9.92153&6.12650&6.24914&6.21484&5.75515&6.13593\\ \hline
Q9H9B1           &ANK$_t$&3.11864&9.67296&6.01194&6.16525&6.17296&5.71148&6.09592\\ \hline
Q9H9E1           &ANK$_t$&2.69169&9.37380&5.88498&6.17252&6.11502&5.69649&6.14058\\ \hline
Q9HBA0           &ANK$_t$&2.91102&9.74971&6.01148&6.14351&6.15270&5.81860&5.98737\\ \hline
Q9WUD2           &ANK$_t$&2.97175&8.33180&5.98817&6.17740&6.17740&5.74901&6.00920\\ \hline
Q9Y5S1           &ANK$_t$&2.95812&9.47709&6.00196&6.20746&6.18521&5.78338&5.99804\\ \hline
Q9Z2X2           &ANK$_t$&2.49567&9.52597&6.06494&6.22078&6.12554&5.70130&6.12121\\ \hline
\rowcolor{black}&&&&&&&&\\ \hline
A5LNI9           &DEH$_t$&2.67899&6.05447&9.83463&6.26459&6.26848&5.81323&6.09339\\ \hline
B0A4S9           &DEH$_t$&2.44253&6.02586&9.64368&6.35920&6.25000&5.57759&5.94540\\ \hline
B2Z3V8           &DEH$_t$&2.33958&5.96458&7.08542&5.98958&6.06875&5.66458&5.94792\\ \hline
B4A833           &DEH$_t$&2.46610&5.99322&8.15085&6.22034&6.22373&5.83051&5.97966\\ \hline
B4ABS1           &DEH$_t$&2.56400&5.89000&9.87400&6.17800&6.18200&5.62600&6.06200\\ \hline
C2JJ26           &DEH$_t$&2.68750&6.12305&9.38672&6.29492&6.21289&5.76367&6.07227\\ \hline
C6IG92           &DEH$_t$&2.05755&6.01439&7.27698&6.15108&6.08633&5.75540&6.00719\\ \hline
C7RF86           &DEH$_t$&2.73298&5.87304&6.76702&6.29974&6.07984&5.74476&5.87827\\ \hline
D4FVT8           &DEH$_t$&2.19787&5.89362&8.07872&6.06809&5.94894&5.82128&5.99574\\ \hline
D4V4M5           &DEH$_t$&2.45287&5.94467&8.49795&6.28893&6.12910&5.76025&5.96516\\ \hline
D7IJ01           &DEH$_t$&2.50581&5.94767&8.51550&6.41279&6.25388&5.69961&6.00969\\ \hline
D8FP15           &DEH$_t$&2.39331&6.10460&7.87657&6.33891&6.21339&5.65272&6.12971\\ \hline
G9RLJ0           &DEH$_t$&2.37218&5.90414&9.27256&6.14850&6.09211&5.82143&5.91165\\ \hline
K2T267           &DEH$_t$&2.35754&6.01117&9.72067&6.43017&6.25140&5.71508&6.16201\\ \hline
M2DKS0           &DEH$_t$&2.65175&6.03113&9.75681&6.29572&6.21012&5.75486&6.11284\\ \hline
M4SEQ9           &DEH$_t$&2.99457&6.07677&9.59375&6.65353&6.29280&5.76562&6.17867\\ \hline
O08575           &DEH$_t$&2.80263&6.06297&9.89098&6.16071&6.13628&5.84680&6.05357\\ \hline
O15305           &DEH$_t$&2.49390&5.89634&9.65854&6.16057&6.13618&5.72154&5.90447\\ \hline
O29777           &DEH$_t$&3.16667&6.24067&8.13184&6.63619&6.45211&5.74192&6.25062\\ \hline
O32125           &DEH$_t$&2.43045&5.90414&9.22556&6.09962&6.15602&5.75376&6.05451\\ \hline
O32220           &DEH$_t$&3.18080&6.11534&9.90337&6.71384&6.35599&5.71135&6.18516\\ \hline
O59346           &DEH$_t$&2.41286&6.04149&8.77801&6.26556&6.19917&5.55602&5.92116\\ \hline
O67920           &DEH$_t$&2.47546&6.22086&7.06135&6.55215&6.36196&5.61350&6.20245\\ \hline
P05023           &DEH$_t$&2.87004&6.05433&9.72981&6.49046&6.17181&5.73862&6.05727\\ \hline
P06685           &DEH$_t$&2.97801&6.00880&9.82551&6.48876&6.20919&5.75269&6.02151\\ \hline
P0AE22           &DEH$_t$&2.35443&6.00000&9.87764&6.17722&6.25316&5.70886&6.11392\\ \hline
P20649           &DEH$_t$&2.97102&6.03267&9.51106&6.55321&6.20864&5.74078&6.03793\\ \hline
P35670           &DEH$_t$&2.25532&6.05674&9.57092&6.34752&6.26241&5.78014&6.26950\\ \hline
P78330           &DEH$_t$&2.36000&6.03111&9.92889&6.24000&6.14222&5.66667&6.05778\\ \hline
P94592           &DEH$_t$&2.41579&6.03158&9.44737&6.24912&6.19298&5.84211&6.07719\\ \hline
Q04656           &DEH$_t$&2.27222&6.25000&8.37778&6.63889&6.09444&5.65000&6.23889\\ \hline
Q11S56           &DEH$_t$&2.44170&5.91519&6.98057&6.18728&6.32155&5.74558&6.03534\\ \hline
Q2T109           &DEH$_t$&2.19519&6.19251&9.71123&6.48663&6.31016&5.84492&6.14439\\ \hline
Q3UGR5           &DEH$_t$&2.57469&6.19087&9.45021&6.40664&6.33195&5.65975&6.09129\\ \hline
Q5EBQ9           &DEH$_t$&2.28538&5.95519&9.73585&6.05896&6.03538&5.75708&5.92217\\ \hline
Q5SJQ3           &DEH$_t$&3.02124&6.41313&7.03282&6.67181&6.52510&5.88803&6.38224\\ \hline
Q60048           &DEH$_t$&2.94304&6.06610&8.47468&6.42194&6.31786&5.69198&6.09564\\ \hline
Q7ADF8           &DEH$_t$&2.45045&6.10135&9.78829&6.43919&6.46622&5.75450&6.18694\\ \hline
Q7WG29           &DEH$_t$&2.36034&6.02235&8.75978&6.27933&6.20670&5.69832&6.08939\\ \hline
Q8K7R3           &DEH$_t$&2.63386&5.99409&9.34252&6.28150&6.24606&5.78543&6.04921\\ \hline
Q8L1N9           &DEH$_t$&2.52353&5.99216&7.90392&6.29804&6.23137&5.78431&6.13333\\ \hline
Q8TBE9           &DEH$_t$&2.51481&6.07963&9.63704&6.23519&6.09444&5.72407&6.10926\\ \hline
Q96X90           &DEH$_t$&2.56364&6.03864&6.50682&6.33864&6.27955&5.73409&6.08864\\ \hline
Q96XE7           &DEH$_t$&2.50971&6.06553&6.53155&6.06553&6.16262&5.76456&5.91019\\ \hline
Q98I56           &DEH$_t$&2.47005&6.05530&8.15207&6.41475&6.29493&5.81106&6.11982\\ \hline
Q9D020           &DEH$_t$&2.54381&5.96073&9.45317&7.96073&6.17523&5.71601&5.98792\\ \hline
Q9JLV6           &DEH$_t$&2.77969&6.06034&9.43774&6.24425&6.18295&5.81897&6.10057\\ \hline
Q9X0Y1           &DEH$_t$&2.78704&6.19676&7.64583&6.45139&6.30787&5.74769&6.26157\\ \hline
T0QBN5           &DEH$_t$&2.48256&5.96705&9.28876&6.40891&6.14922&5.68798&5.93605\\ \hline
U8H1V1           &DEH$_t$&2.38293&5.98537&9.68780&6.21951&6.20976&5.63902&6.10732\\ \hline
\rowcolor{black}&&&&&&&&\\ \hline
B1MJ53           &HET$_t$&2.57143&6.06656&6.15747&6.45942&6.14123&5.73539&6.07955\\ \hline
D3H0F7           &HET$_t$&2.73485&6.06439&6.14520&6.29672&6.24621&5.70076&5.96843\\ \hline
E6Z0R3           &HET$_t$&2.85146&6.09519&6.01151&6.20607&6.15586&5.75418&6.03033\\ \hline
F4AR88           &HET$_t$&2.58306&6.05980&6.02990&6.98339&6.20266&5.76744&6.08970\\ \hline
O06961           &HET$_t$&2.72886&6.03109&6.10323&6.95771&6.21517&5.74751&6.07587\\ \hline
O52806           &HET$_t$&2.35366&5.92195&6.07805&6.36098&6.24878&5.75122&6.00000\\ \hline
O58456           &HET$_t$&2.52453&6.04528&6.13585&6.24151&6.29057&5.75472&6.04151\\ \hline
P00437           &HET$_t$&2.26569&5.74059&5.76569&6.10460&5.92469&5.77406&5.90795\\ \hline
P00693           &HET$_t$&2.74658&5.95548&6.01256&6.29795&6.15183&5.78425&6.03767\\ \hline
P00772           &HET$_t$&2.46617&5.89286&5.93421&6.17105&6.08083&5.71992&5.91165\\ \hline
P00800           &HET$_t$&2.79197&6.02281&5.94982&6.20529&6.19252&5.78376&6.08120\\ \hline
P00918           &HET$_t$&2.41346&5.88654&5.88269&6.07500&6.06731&5.64423&5.93654\\ \hline
P02883           &HET$_t$&2.25604&5.78261&5.78744&5.97101&5.93237&5.71014&5.80676\\ \hline
P09211           &HET$_t$&2.49524&6.02143&5.96905&6.27381&6.28810&5.75476&5.97381\\ \hline
P0A6C8           &HET$_t$&2.61822&6.09884&6.33140&6.66473&6.37791&5.65310&6.19961\\ \hline
P0A8M3           &HET$_t$&2.70093&5.90576&5.89486&6.37305&6.00701&5.76090&5.80763\\ \hline
P0C0Y9           &HET$_t$&2.53571&5.91396&5.93019&7.41234&6.19318&5.64448&6.06981\\ \hline
P0C512           &HET$_t$&2.67610&5.88050&6.00000&6.87002&6.17820&5.76730&6.01468\\ \hline
P23472           &HET$_t$&2.58360&6.07395&5.99035&6.18650&6.12219&5.67846&5.96463\\ \hline
P23904           &HET$_t$&2.32700&5.78059&5.80591&5.98734&5.93249&5.68354&5.89873\\ \hline
P26663           &HET$_t$&3.44219&6.05797&6.01412&6.59169&6.19950&5.76262&6.04336\\ \hline
P27448           &HET$_t$&3.01394&6.09429&5.99734&6.62351&6.18327&5.74635&6.13944\\ \hline
P29476           &HET$_t$&3.05353&5.96011&5.89923&6.22113&6.07628&5.77327&5.94822\\ \hline
P32169           &HET$_t$&2.43431&5.97263&5.83029&6.18431&6.17701&5.79745&5.93248\\ \hline
P35202           &HET$_t$&2.46865&5.98433&5.95611&6.15361&6.10031&5.68025&5.89655\\ \hline
P37352           &HET$_t$&2.85831&5.93677&6.14286&6.27166&6.19906&5.77986&6.03747\\ \hline
P46154           &HET$_t$&2.68296&6.07519&6.07769&6.62155&6.29825&5.74937&6.07519\\ \hline
P50586           &HET$_t$&2.77129&6.20594&6.07723&6.32079&6.21782&5.68713&6.20000\\ \hline
P61086           &HET$_t$&2.24500&6.16750&6.14250&6.61250&6.22750&5.76750&6.05750\\ \hline
P69834           &HET$_t$&2.49338&6.04470&6.11424&6.30960&6.18377&5.79305&6.14735\\ \hline
Q26997           &HET$_t$&2.51522&5.87174&5.86739&6.20217&6.11087&5.68913&5.92826\\ \hline
Q3IWB0           &HET$_t$&3.08891&6.21797&6.26960&6.52008&6.36520&5.67113&6.29446\\ \hline
Q51723           &HET$_t$&2.68750&5.94174&5.96716&6.24470&6.07309&5.69809&5.82945\\ \hline
Q54727           &HET$_t$&2.93687&6.01435&6.01148&6.13343&6.14347&5.74892&6.02296\\ \hline
Q5TA50           &HET$_t$&2.29907&6.06776&6.09579&6.21729&6.24065&5.68925&6.10047\\ \hline
Q5TLG6           &HET$_t$&2.25446&5.71652&5.75223&5.94866&5.90848&5.70313&5.78795\\ \hline
Q6DLV0           &HET$_t$&3.47434&5.90398&5.95619&6.16062&6.14145&5.78451&5.98982\\ \hline
Q6G441           &HET$_t$&2.62381&6.08690&6.14405&6.73571&6.28929&5.73214&6.14881\\ \hline
Q70C53           &HET$_t$&2.64941&5.95118&5.86538&6.02515&6.10207&5.67012&5.83284\\ \hline
Q873X9           &HET$_t$&2.76790&6.00924&5.93533&6.25173&6.10855&5.70439&6.18476\\ \hline
Q8A7T5           &HET$_t$&2.70487&6.09939&5.96552&6.42394&6.12779&5.68154&6.00000\\ \hline
Q8DCF5           &HET$_t$&2.55389&5.94461&5.98653&6.62575&6.18413&5.71407&5.94461\\ \hline
Q8TX37           &HET$_t$&2.88268&6.18575&6.27793&6.38128&6.32542&5.62430&6.19693\\ \hline
Q97DM1           &HET$_t$&2.62946&5.79737&5.70544&5.97373&5.93809&5.81426&5.92120\\ \hline
Q97VM5           &HET$_t$&2.67969&6.12370&6.19141&6.36849&6.36328&5.81641&6.19401\\ \hline
Q9IFX1           &HET$_t$&2.94501&5.97251&5.90729&6.18734&6.15921&5.75000&6.05946\\ \hline
Q9KFI6           &HET$_t$&2.55152&5.92974&6.00234&6.08665&6.00937&5.61827&5.89930\\ \hline
Q9NUI1           &HET$_t$&2.67123&6.12842&6.03253&6.69007&6.35788&5.73801&6.25514\\ \hline
Q9P286           &HET$_t$&2.90334&6.14812&5.92629&6.34075&6.16690&5.78999&6.11822\\ \hline
Q9WZY5           &HET$_t$&2.49187&5.90041&5.88415&6.27439&6.13211&5.68902&5.93699\\ \hline
\rowcolor{black}&&&&&&&&\\ \hline
A6ZU46            &WD$_t$&2.98594&6.10686&5.99100&6.26434&6.25984&5.77278&8.46175\\ \hline
O14727            &WD$_t$&3.13301&6.02123&5.91306&6.11098&6.12460&5.78966&9.77244\\ \hline
O24456            &WD$_t$&3.03211&6.03364&6.06422&6.15596&6.21713&5.79511&8.88838\\ \hline
O75530            &WD$_t$&2.50227&5.92971&6.24490&6.08617&6.06576&5.68027&9.93991\\ \hline
O76071            &WD$_t$&3.14159&6.00885&5.82006&6.06490&6.01180&5.72271&9.83776\\ \hline
O88879            &WD$_t$&3.12770&6.09848&5.94396&6.15452&6.11769&5.78863&8.69936\\ \hline
O89053            &WD$_t$&2.79501&6.10629&5.96746&6.15401&6.13883&5.78091&9.33731\\ \hline
P07834            &WD$_t$&3.17522&6.12195&6.09114&6.24390&6.15148&5.76252&7.62323\\ \hline
P16649            &WD$_t$&3.38079&6.47546&6.20477&6.45512&6.27630&5.71248&8.45231\\ \hline
P26449            &WD$_t$&2.56452&5.95894&5.93255&6.08211&6.14076&5.81525&6.51026\\ \hline
P36037            &WD$_t$&2.89441&6.06713&6.01958&6.16783&6.20140&5.80420&7.57832\\ \hline
P38011            &WD$_t$&2.86991&6.05016&5.88088&6.19436&6.18182&5.68966&9.00940\\ \hline
P38262            &WD$_t$&2.84019&6.04860&6.01682&6.23178&6.18505&5.76636&7.55140\\ \hline
P38968            &WD$_t$&3.37510&6.17203&5.97486&6.25373&6.22545&5.77219&7.29929\\ \hline
P40217            &WD$_t$&2.67867&5.92507&5.75216&6.10951&6.08069&5.70317&9.53458\\ \hline
P46680            &WD$_t$&2.88049&6.02602&5.94959&6.09268&6.10081&5.67805&8.23171\\ \hline
P53011            &WD$_t$&2.55444&6.04298&5.82521&6.03438&6.07450&5.71347&9.21490\\ \hline
P53196            &WD$_t$&2.57434&5.94245&5.94724&6.20144&6.09592&5.76739&6.54197\\ \hline
P54311            &WD$_t$&2.77647&6.03088&5.85735&6.12500&6.07794&5.80147&9.96471\\ \hline
P55735            &WD$_t$&2.75155&5.86180&5.79037&5.96739&5.98602&5.65683&9.56832\\ \hline
P61964            &WD$_t$&3.37575&6.11826&5.95359&6.08832&6.16916&5.80090&9.97305\\ \hline
P61965            &WD$_t$&3.37575&6.11826&5.95359&6.08832&6.16916&5.80090&9.97305\\ \hline
P62871            &WD$_t$&2.77647&6.03088&5.85735&6.12500&6.07794&5.80147&9.96471\\ \hline
P62881            &WD$_t$&2.69747&5.92911&5.86329&6.13165&6.00506&5.84810&9.76709\\ \hline
P63005            &WD$_t$&3.07561&5.93293&5.89878&6.03293&6.03537&5.62073&9.96463\\ \hline
P63244            &WD$_t$&2.84700&5.94322&5.83912&6.06625&6.05994&5.64984&9.91167\\ \hline
P78406            &WD$_t$&2.50272&5.78668&5.71060&5.96603&6.04212&5.71875&9.93071\\ \hline
P78774            &WD$_t$&2.69629&6.02653&5.93899&6.09019&6.14854&5.75862&7.20424\\ \hline
P78972            &WD$_t$&2.80533&6.03381&5.89037&6.16906&6.09324&5.69160&7.39857\\ \hline
Q02793            &WD$_t$&2.66877&5.98741&5.91688&6.15113&6.07809&5.74307&6.59950\\ \hline
Q03774            &WD$_t$&2.73198&5.98311&5.94257&6.05743&6.14752&5.76014&7.34797\\ \hline
Q04491            &WD$_t$&2.68013&5.89226&5.90236&6.06397&6.09764&5.72054&9.52862\\ \hline
Q04724            &WD$_t$&2.96234&6.08117&5.96299&6.21623&6.11753&5.78896&9.80065\\ \hline
Q05583            &WD$_t$&2.69697&6.08939&5.89848&6.18333&6.06515&5.67727&7.92121\\ \hline
Q09028            &WD$_t$&2.80118&6.04471&5.98118&6.01647&6.04941&5.77647&9.89412\\ \hline
Q11176            &WD$_t$&2.95172&5.98691&5.91326&6.09820&6.20131&5.77578&8.57201\\ \hline
Q13216            &WD$_t$&2.69192&5.89268&5.90530&6.03409&6.05934&5.72601&9.45455\\ \hline
Q16576            &WD$_t$&2.65059&6.02118&5.98824&5.99765&5.98588&5.83765&9.82588\\ \hline
Q24572            &WD$_t$&2.70000&5.99651&5.96628&5.99884&5.97791&5.76628&9.45698\\ \hline
Q24D42            &WD$_t$&2.53499&5.81924&5.79883&6.01458&6.08455&5.78134&8.55685\\ \hline
Q2YDS1            &WD$_t$&2.79839&6.00504&5.90423&6.18044&6.17843&5.75504&7.13407\\ \hline
Q58CQ2            &WD$_t$&2.57527&5.98790&5.91532&6.06855&6.11694&5.72177&9.56452\\ \hline
Q6CN23            &WD$_t$&2.57965&5.80531&5.88201&5.96755&5.98525&5.76696&6.28614\\ \hline
Q8LNY6            &WD$_t$&2.65526&5.90395&5.88553&6.00132&6.06974&5.71711&7.77500\\ \hline
Q921E6            &WD$_t$&2.50227&5.92971&6.24490&6.08617&6.06576&5.68027&9.93991\\ \hline
Q92466            &WD$_t$&2.57260&5.96487&5.90867&6.07728&6.10304&5.68852&8.86768\\ \hline
Q969H0            &WD$_t$&3.17751&6.15842&6.00424&6.24470&6.13296&5.71994&9.45686\\ \hline
Q96MX6            &WD$_t$&2.56162&6.00280&5.80392&6.02521&6.08683&5.73389&9.85154\\ \hline
Q9GZS3            &WD$_t$&2.86721&5.94098&6.00328&6.17705&6.08525&5.70492&9.92951\\ \hline
Q9Y297            &WD$_t$&2.90992&6.00000&5.92066&6.06446&6.04463&5.71240&9.96446\\ \hline
\end{longtable}
\clearpage
\newpage



%
%
%

%
%

\end{document}